\documentclass[prb,aps,twocolumn,showpacs,amsmath,amssymb]{revtex4-1}

\usepackage{graphicx}
\usepackage{amssymb}
\usepackage{psfrag}
\usepackage{epsfig}
\usepackage{color}
\usepackage[tight]{subfigure}
\definecolor{dred}{rgb}{0.7,0.0,0.0}
\usepackage{dcolumn}
\usepackage{bm}
\allowdisplaybreaks 
\usepackage{braket}
\usepackage[dvipdfmx]{hyperref}
\hypersetup{colorlinks=true, linkcolor=blue, citecolor=blue, filecolor=blue, urlcolor=blue, pdftitle=, pdfauthor=, pdfsubject=, pdfkeywords=}

\newcommand{\cnod}{c^{\phantom{\dagger}}}
\newcommand{\cdag}{c^\dagger}

\begin{document}

\title{Fractional Chern insulator on a triangular lattice of strongly correlated $t_{2g}$ electrons}
\author{Stefanos Kourtis}
\author{J\"orn  W.F. Venderbos}
\author{Maria Daghofer}
\affiliation{Institute for Theoretical Solid State Physics, IFW Dresden, 01171 Dresden, Germany}
\date{\today}

\begin{abstract}
We discuss the low-energy limit of three-orbital Kondo-lattice and
Hubbard models describing $t_{2g}$ orbitals on a triangular lattice 
near half-filling. We analyze how very flat single-particle bands with non-trivial
topological character, a Chern number $C=\pm1$, arise both in the limit of
infinite on-site interactions as well as in more realistic
regimes. Exact diagonalization is then used to investigate
an effective one-orbital spinless-fermion model at fractional fillings including
nearest-neighbor interaction $V$; it reveals signatures of fractional
Chern insulator (FCI) states for several filling fractions. In addition to
indications based on energies, e.g. flux insertion and
fractional statistics of quasiholes, Chern numbers are obtained. It is
shown that FCI states are robust against disorder in the underlying magnetic
texture that defines the topological character of the  band. We
also investigate competition between a FCI state and a 
charge density wave (CDW) and discuss the effects of particle-hole asymmetry and
Fermi-surface nesting. FCI states turn out to be rather robust and do
not require very flat bands, but can also arise when filling or an
absence of Fermi-surface nesting disfavor the competing
CDW. Nevertheless, very flat bands allow FCI states to be induced by
weaker interactions than those needed for more dispersive bands.  
\end{abstract}

\maketitle

\section{Introduction} \label{sec:intro}

Quantum liquids are among the most sought after states of matter. One
celebrated class of such quantum liquids, the fractional quantum-Hall
(FQH) states,~\cite{Laughlin1983,Haldane1983,Halperin1984} has been a
locus of scientific attention for almost three decades. Initially
introduced to explain the fractional quantum-Hall effect (FQHE)
observed in semiconductor devices,~\cite{Tsui1982} they describe
interacting electrons constrained in two dimensions and subject to a
strong perpendicular magnetic field. Fractionally filling a magnetic
Landau level (LL) then yields an incompressible FQH liquid and gives rise
to a precisely quantized Hall conductivity. In addition, the
quasiparticles of these states obey anyonic
statistics,~\cite{Wilczek1982,Haldane1991} which can be either Abelian
or non-Abelian, the latter fulfilling an essential condition for
fault-tolerant quantum computation.~\cite{Nayak2008} 


An alternative route to quantum Hall liquids is via tightly bound
electrons moving in a magnetic texture. The simplest such magnetic
texture is a uniform magnetic field perpendicular to the plane of the
system. For non-interacting, tightly bound electrons, this gives rise
to an integer quantum-Hall (IQH) state and leads to Hofstadter's
fractal energy spectrum,~\cite{Hofstadter1976} whereas in a
fractionally filled tight-binding model of electrons interacting via a
screened Coulomb repulsion, the realization of a FQH state is
possible.~\cite{Kliros1991} Haldane noted~\cite{haldane1988} that an external
magnetic field is not the only viable path to IQH states and introduced a 
honeycomb-lattice model, with complex hoppings that break
time-reversal invariance, and defined the first integer Chern insulator
(CI), in which the total magnetic field through the unit cell averages to
zero. In the meantime, mechanisms leading to complex
hoppings with the necessary properties have been identified. One realization may
be found in strongly spin-orbit coupled semiconductor materials that
are ferromagnetically ordered.~\cite{Qi:2006jm,Liu:2008ev} Another
possibility arises through the coupling of itinerant electrons to localized magnetic
moments,~\cite{Ohgushi2000} for example the Kondo-lattice model on the
triangular lattice supports a non-trivial magnetic texture, which
induces an integer-quantized Hall conductivity of the itinerant
electrons.~\cite{Martin2008}  

A considerable body of recent research has addressed the question
whether the lattice counterpart of the FQHE can be observed when topologically non-trivial
bands, called Chern bands, are fractionally filled and electrons are 
interacting. Several numerical studies using exact diagonalization
techniques have convincingly established the existence of
Laughlin-series states in an number of different models on various
lattices.~\cite{Tang2011,Sun2011,Neupert2011,Sheng2011,Wang2011,Regnault2011,Wu2012} These
systems have since been called fractional Chern insulators (FCI). Very
recent work has reported FCI states beyond Laughlin
fractions.~\cite{Venderbos2012,Liu2012a,2012arXiv1207.6094L} In
addition to reproducing the known FQHE on a lattice and potentially at
higher temperatures, CIs also offer the intriguing possibility of a Chern number higher than one, a departure from the analogy
with LLs. Recent studies have explored this direction by constructing
models that have higher Chern numbers and studying possible FCI
states.~\cite{Trescher2012,Yang2012a,Wang2012,Liu2012b,Sterdyniak2012,2012arXiv1207.4097G} From
the analytical side, the problem of FCI states has been approached by a
careful study of emergent translational symmetries~\cite{Bernevig2012}
and many-body trial wave
functions.~\cite{Qi2011,Wu2012b,Lee2012} Others have examined the
algebraic properties of the density operators projected onto one Chern
band and made a comparison with the Girvin-MacDonald-Platzman algebra
that is satisfied by the lowest LL density operators in the continuum
FQHE.~\cite{Bernevig2012,Murthy2012,Murthy2011,Parameswaran2012,Goerbig2012} 


In numerical studies of FCI states, one typically starts by adding inter-site
interactions to topologically nontrivial but non-interacting CI
models. As potential realizations of the non-interacting and nearly
flat ``parent'' bands, cold atoms,~\cite{Sun2011} oxide
heterostructures,~\cite{Xiao:2011} strained
graphene,~\cite{PhysRevLett.108.266801} and strongly correlated
multi-orbital models for layered
oxides~\cite{Venderbos2011,Venderbos2012} have been proposed. In the
present paper, we build upon this last approach and thus focus our attention on a strongly
correlated model on the triangular lattice. It was shown that on the
mean-field level and near half-filling, a magnetically ordered CI 
with a very flat single-particle band emerges.  Doping this nearly flat band to
fractional fillings was shown to give rise to FCI states within the
framework of an effective single-orbital model. Here, we give
a detailed account of the mapping onto the effective single-orbital model and
show how such topologically nontrivial and nearly flat bands emerge 
in Kondo-lattice and Hubbard models both for the limit of
infinite onsite interactions and for more realistic intermediate
interaction strength. 

We then provide extensive numerical evidence, based on both eigenvalue spectra (e.g., ground-state degeneracy and spectral flow) and eigenstate properties (e.g., many-body Berry curvature and Hall conductivity), for the existence of
FCI states in this model by using exact diagonalization. We discuss the
robustness of FCI states against disorder originating from single-site
defects in the magnetic ordering, which will always occur in realistic
situations. Another very relevant issue in the context of FCI states
is their competition with other
phases,~\cite{2012arXiv1207.4097G,2012arXiv1207.6094L} e.g.
symmetry-broken states such as a 
charge-density wave (CDW). We study this issue by using a method that
does not project onto the nearly flat Chern band and thus keeps the
effects of dispersion. The model considered in this work allows
for a careful study of the competition between finite dispersion and
interactions. We map out a phase diagram for filling fractions
$\nu=1/3$ and $\nu=2/3$. In the latter case, filling permits a commensurate
CDW, whereas in the former it does not. The CDW is favored by Fermi-surface (FS) nesting, and we accordingly find the FCI at $\nu=2/3$ to be far more stable when the bands are poorly nested.

This paper is organized as follows: in Sec.~\ref{sec:kondo_hubb}, we
extend the discussion of flat and topologically nontrivial bands
arising in $t_{2g}$ models and derive the effective model.
Numerical results on the model are presented in Sec.~\ref{sec:results},
where we first focus on information obtained from eigenenergies,
namely gaps and flux insertion (Sec.~\ref{sec:spectrum}) and fractional statistics of
charged excitations in FCI states based on a recently introduced state-counting
argument (Sec.~\ref{sec:stat}).~\cite{Bernevig2012} We then add
information obtained from the eigenstates: the Hall conductivity, which
allows us to address the impact of impurities in
Sec.~\ref{sec:disorder} and the static charge-structure factor, which
allows us to discuss the competition with the CDW at filling
$\nu=2/3$ in Sec.~\ref{sec:phdiag}.
%
We conclude with some remarks
summarizing the main points in Sec.~\ref{sec:conclusions}

\section{Topologically nontrivial and nearly flat bands
  in strongly correlated multi-orbital systems}\label{sec:kondo_hubb}

\begin{figure}
 \centering
\subfigure[]{\includegraphics[width=0.4\columnwidth]{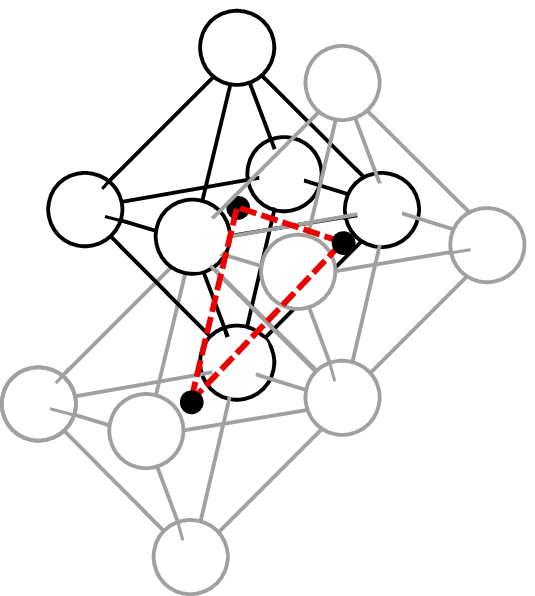}\label{fig:oct}}
\subfigure[]{\includegraphics[width=0.5\columnwidth]{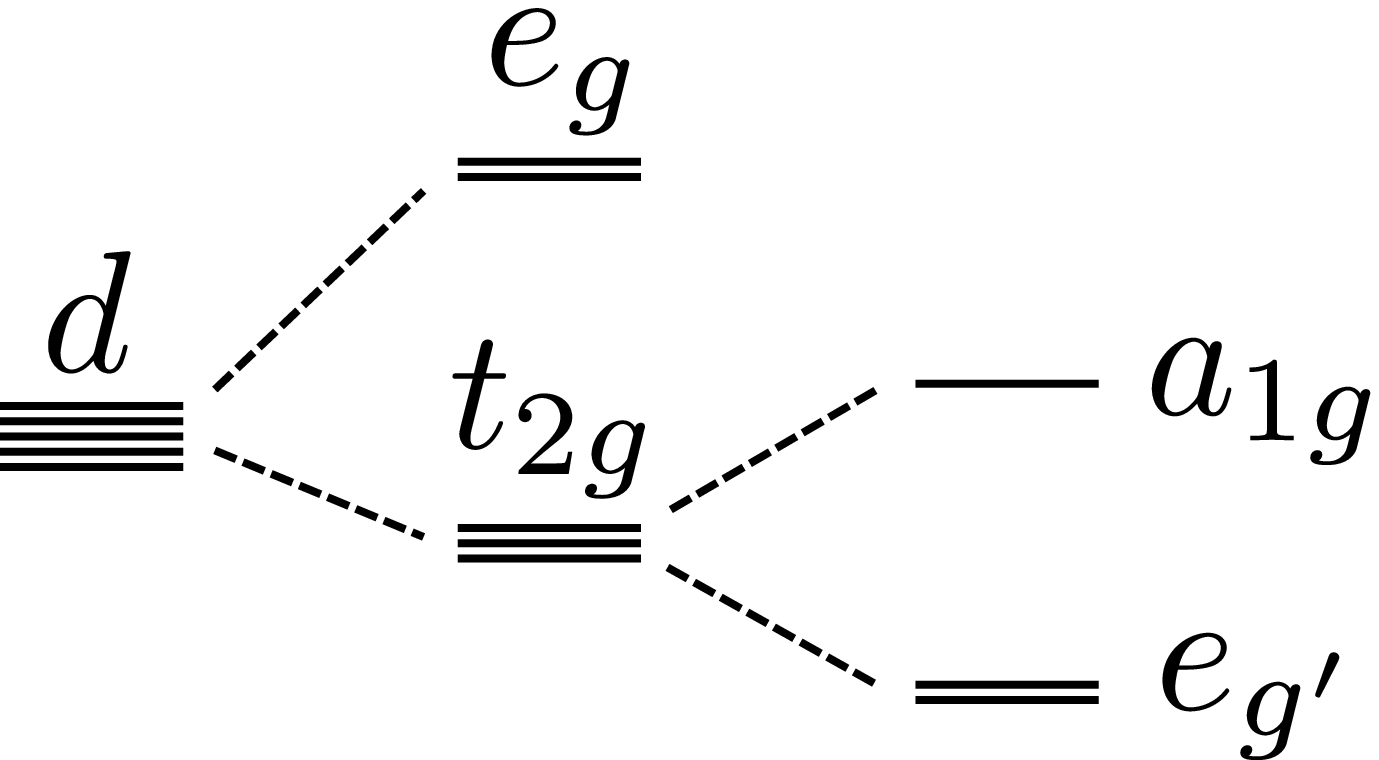}\label{fig:split}}
\caption{(Color online) (a) Illustration of a triangular-lattice plane
  built of edge-sharing oxygen octahedra. (b) The five $d$ orbitals of the
  transition-metal ion in the center are split into an $e_g$ doublet
  and a $t_{2g}$ triplet due to the local cubic symmetry; the latter
  is further split into one $a_{1g}$ state and an $e_g'$ doublet. (The
splitting between the latter is exaggerated here for visibility.)}\label{fig:oct_orb}
\end{figure}

In Ref.~\onlinecite{Venderbos2011}, it was shown that both $e_g$ and
$t_{2g}$ orbital manifolds in octahedral coordination can reduce the bandwidth of topologically
nontrivial bands. For a schematic illustration of 
orbital degeneracy in $d$-electron systems see Fig.~\ref{fig:oct_orb}.
In particular, this was discussed for the
spin-chiral phase arising in Kondo-lattice
models on the triangular lattice\cite{Martin2008,Akagi:2010p083711,Kumar:2010p216405,Kato_FKLM_tri_2010}
at quarter and three-quarter fillings. 
While the flat band of interest mixes both the
$3z^2-r^2$ and the $x^2-y^2$ orbitals in the $e_g$ case, it is
dominated by a particular orbital state in the
$t_{2g}$ manifold, the $a_{1g}$ state, which allowed the straightforward mapping onto an
effective one-band model.~\cite{Venderbos2012} In this section, we
first review the basic setup of
the orbital symmetries in Sec.~\ref{sec:orbs} and the modifications of the hopping through
the magnetic order in Sec.~\ref{sec:berry}, as we believe that it may be of value to the
reader. After this presentation of the physical ingredients at work
here, we discuss how the nearly flat $a_{1g}$ band arises due to effective
longer-range hoppings. After the simplest scenario of infinite Hund's rule coupling in
Sec.~\ref{sec:Jinf}, we go to the more realistic
multi-orbital model with finite interactions in
Sec.~\ref{sec:Jfinite}, and finally decide on a relatively simple
effective model in Sec.~\ref{sec:eff_model}, which nevertheless
captures several important features.  

\subsection{Impact of orbital symmetries on the one-particle Hamiltonian} \label{sec:orbs}

In many transition-metal (TM) compounds, the local symmetry around a
TM ion is cubic, with ligand oxygens forming 
an octahedron, as depicted in Fig.~\ref{fig:oct}. This
splits the degeneracy between the $d$ levels, because the two $e_g$
orbitals point toward the negatively charged oxygens, while the three
$t_{2g}$ levels have their lobes in between. Consequently, the energy
of $e_g$ levels is higher. Depending on the total electron filling,
the valence states may be found in either manifold. We are here
discussing the situation where the three $t_{2g}$ levels share $2.5$
to $3$ electrons and the $e_g$ levels are empty. Furthermore, we
consider the case of a layered triangular lattice, as can be realized
in compounds of the form ABO$_2$.

In this geometry, the octahedra are edge sharing  and electrons (or
holes) can hop from one TM ion to its neighbor either through direct
overlap or via the ligand oxygens. The hopping symmetries can be most
easily worked out using the usual basis functions for the $t_{2g}$
states, $|xz\rangle$, $|yz\rangle$, and
$|xy\rangle$~\cite{Pen97,Koshibae03}  and following Refs.~\onlinecite{har80,slater}.
Considering hopping for bonds along the ${\bf a}_1$ direction and
choosing the local coordinate system such that this corresponds to the
$(1,1)$ direction in the $x$-$y$ plane, one
finds that direct hopping $t_d$ is only relevant for the $xy$
orbital and conserves orbital flavor. Due to the $90^\circ$ angle of
the TM-O-TM bond, oxygen-mediated hopping $t_0$ is, on the other hand, 
mostly via the oxygen-$p_z$ orbital and mediates processes between
$xz$ and $yz$  states, thereby always \emph{changing} orbital
flavor. Hoppings along the other two, symmetry-related, directions
${\bf a}_2$ and ${\bf a}_3$ are obtained by symmetry transformations.

These hoppings can then be  expressed in orbital- and direction-
dependent matrix elements $t_{{\bf a}_k}^{\alpha,\beta}$, where $\alpha$ and $\beta$
denote orbitals ($xz$, $yz$, and $xy$) and ${\bf a}_k$ the
direction. They are given by
\begin{gather}
 \hat{T}_{{\bf a}_1} =  \begin{pmatrix}
t_{dd} & 0 & 0\\
0  & 0 & t_0 \\
0  & t_0 & 0 \\
\end{pmatrix}, \ 
\hat{T}_{{\bf a}_2} = \begin{pmatrix}
0  & 0 & t_0\\
0  & t_{dd}& 0 \\
t_0  & 0 & 0 \\
\end{pmatrix},\nonumber \\  
\hat{T}_{{\bf a}_3} =  \begin{pmatrix}
0  & t_0 & 0\\
t_0  & 0 & 0 \\
0  & 0 & t_{dd}\\
\end{pmatrix}\label{eq:hopp_xyz} 
\end{gather}
for NN bonds along the three directions ${\bf a}_1$, ${\bf a}_2$,
${\bf a}_3$. The two hopping processes are expected to be of comparable strength,
but with $|t_d|\lesssim |t_0|$ for $3d$ elements, and will typically have opposite sign.~\cite{Koshibae03}

If the width of a triangular layer made
of octahedra is compressed (extended), the energy of the highly symmetric orbital
state $|a_{1g}\rangle=(|xz\rangle+|yz\rangle+|xy\rangle)/\sqrt{3}$  is
raised (lowered) with respect to the remaining orbital doublet
($e_g'$), see Fig.~\ref{fig:split} for illustration. This energy shift
can be written as  
\begin{equation}\label{eq:JT_xyz}
H_{\textrm{JT}} =  -E_{\textrm{JT}} (n_{e_{g+}}+n_{e_{g-}}
-2n_{a_{1g}})/3
\end{equation}
and depends on the Jahn-Teller effect as well as on the
lattice.~\cite{Koshibae03} Especially for large splitting between
$a_{1g}$ and $e_g'$ states, which may also be enhanced through onsite
Coulomb interactions, see Sec.~\ref{sec:Jfinite}, it is more appropriate to use a basis that
reflects the triangular lattice symmetry. We thus go over into the $(a_{1g},e_{g,1}',e_{g,2}')$ basis, which is
done via~\cite{Koshibae03} 
\begin{equation}
\left(\begin{array}{c} a_{1g}\\e_{g,1}'\\e_{g,2}'
\end{array} \right) = \hat{U} \left(\begin{array}{c} xz\\yz\\xy
\end{array} \right) =  \frac{1}{\sqrt{3}}\begin{pmatrix}
1 & 1 & 1 \\
1& e^{ i 2\pi/3}& e^{ -i2\pi/3} \\
1&  e^{i 4\pi/3}& e^{- i4\pi/3}
\end{pmatrix} \left(\begin{array}{c} xz\\yz\\xy
\end{array} \right).
\end{equation}
The transformed hopping matrices $\widetilde{T}_{{\bf a}_i}$ are then obtained from Eq.~(\ref{eq:hopp_xyz}) as 
\begin{gather}
\widetilde{T}_{{\bf a}_1} = \hat{U}^{\dagger}\hat{T}_{{\bf a}_1} \hat{U} = \frac{1}{3} \begin{pmatrix}
3t_0 +\delta t& \delta t  & \delta t  \\
\delta t  & \delta t & 3t_0 +\delta t \\
\delta t & 3t_0 +\delta t & \delta t
\end{pmatrix}, \nonumber\\  
\widetilde{T}_{{\bf a}_2} =\hat{U}^{\dagger}\hat{T}_{{\bf a}_2} \hat{U}  = \frac{1}{3}\begin{pmatrix}
3t_0 +\delta t& \delta t \,\omega& \delta t\, \omega^{-1} \\
\delta t \, \omega^{-1} & \delta t & (3t_0 +\delta t)\omega \\
\delta t  \,\omega& (3t_0 +\delta t)\omega^{-1} & \delta t
\end{pmatrix}, \nonumber \\
\widetilde{T}_{{\bf a}_3} =\hat{U}^{\dagger}\hat{T}_{{\bf a}_3} \hat{U} = \frac{1}{3} \begin{pmatrix}
3t_0 +\delta t& \delta t \,\omega^{-1} & \delta t \,\omega \\
\delta t \,\omega & \delta t & (3t_0 +\delta t)\omega^{-1} \\
\delta t \,\omega^{-1}& (3t_0 +\delta t)\omega & \delta t
\end{pmatrix},\label{eq:hopp_a1g_eg} 
\end{gather}
where $\delta t = t_{dd}-t_0$ and $\omega=e^{i2\pi/3}$. Observe that
the intra-orbital hopping of the $a_{1g}$ state is the same in all
three lattice directions, as expected for $a_{1g}$ symmetry. However,
we also see that hopping elements mix all three orbitals. 

\subsection{Hopping topology through magnetic order: Berry phases} \label{sec:berry}

\begin{figure}[t!]
 \centering
\includegraphics[width=\columnwidth]{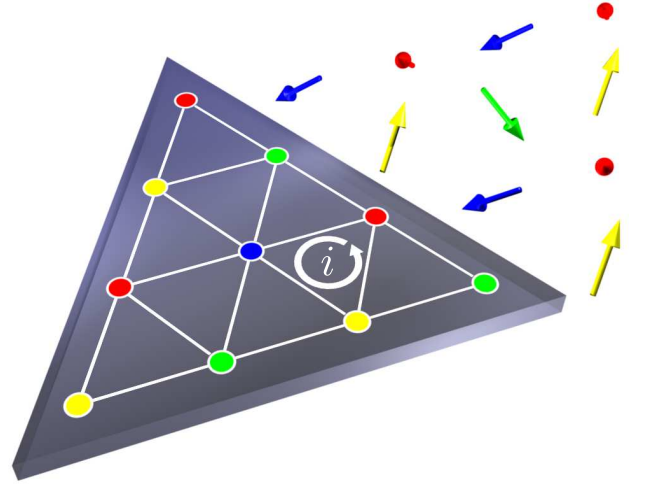}
\caption{(Color online) Illustration of chiral spin pattern and effective model
  including the magnetic texture as phase factors in hoppings. }\label{fig:chiral}
\end{figure}

In addition to an orbital degree of freedom, we now consider coupling
to a localized spin ${\bf S}_{\bf i}$, modelled by a Kondo-lattice model, where the kinetic energy
is given by hopping elements $t_{{\bf a}_k}^{\alpha,\beta}$ taken from
the matrices Eq.~(\ref{eq:hopp_xyz}) or
Eq.~(\ref{eq:hopp_a1g_eg}). This situation is described by 
\begin{align}
\mathcal{H} = \sum_{\stackrel{\langle i,j\rangle\parallel {\bf a}_k,,\sigma}{{\bf a}_k,\alpha,\beta}} 
t_{{\bf a}_k}^{\alpha,\beta} \cdag_{i,\sigma,\alpha} \cnod_{j,\sigma,\beta} 
-J_{\textrm{Kondo}}\sum_{{\bf i},\alpha} {\bf S}_{\bf i}\cdot{\bf s}_{{\bf i},\alpha}\;
\end{align}
where $\langle i,j\rangle\parallel {\bf a}_k$ denotes NN bonds along
the three directions ${\bf a}_k$, $\alpha$ and $\beta$ are orbital indices,
$\cnod_{i,\sigma,\alpha}$  ($\cdag_{i,\sigma,\alpha} $) annihilates
(creates) an electron with spin $\sigma$ in orbital $\alpha$ at site
$i$, and ${\bf s}_{{\bf i},\alpha}$ is the corresponding vector of
orbital electronic spin operators. $J_{\textrm{Kondo}}$ couples the
itinerant electrons to a generic localized spin ${\bf S}_{\bf i}$, the origin
of which is left unspecified for the moment, but will be discussed extensively later. (It will turn out to be the
spin degree of freedom of the $t_{2g}$ electrons themselves, as in Ref.~\onlinecite{Venderbos2012}.) The
coupling is assumed to be FM, as one would expect from Hund's-rule
coupling. However, we are furthermore going to consider ${\bf S}_{\bf
  i}$ as a \emph{classical} spin, in which case AFM coupling to ${\bf S}_{\bf i}$ would
lead to the equivalent results. 

For classical spins and large $J_{\textrm{Kondo}}$, it is convenient
to go over to a \emph{local} spin-quantization axis, where
``$\uparrow$'' (`$\downarrow$'') refers to parallel (antiparallel) orientation of the electron's spin to the local
axis. This simplifies the Kondo term to 
\begin{equation}\label{eq:kondo_loc}
H_{\textrm{Kondo}} = -J_{\textrm{Kondo}}\sum_{{\bf i},\alpha} {\bf S}_{\bf i}\cdot{\bf
  s}_{{\bf i},\alpha} = -J_{\textrm{Kondo}}\sum_{{\bf i},\alpha}
(n_{\alpha}^{\uparrow} - n_{\alpha}^{\downarrow})/2\;,
\end{equation}
where $n_{\alpha}^{\uparrow}$ ($n_{\alpha}^{\downarrow}$) is the electron
density at site ${\bf i}$ in orbital $\alpha$ with spin (anti-) parallel to
the localized spin.
This local spin definition is particularly convenient when going to
the limit of large $J_{\textrm{Kondo}}$, where one immediately finds
the low-energy states as given by only ``$\uparrow$'' electrons.

On the other hand, the fact that the spin-quantization axis is not the
same at all sites implies that the hopping no longer conserves the new
spin. Instead, hopping acquires as spin-dependent factor
$t_{i,j}^{\alpha,\beta} \to t_{i,j}^{\alpha,\beta,\sigma,\sigma'}
= t_{i,j}^{\alpha,\beta}u_{i,j}^{\sigma,\sigma'}$,~\cite{Dagotto:Book} with
\begin{align}\label{eq:berry_uu}
u^{\uparrow\uparrow}_{ij} &=c_ic_j + s_is_je^{-i(\phi_i-\phi_j)},\\ 
u^{\downarrow\downarrow}_{ij} &= c_ic_j + s_is_je^{i(\phi_i-\phi_j)},\\ 
u^{\sigma\bar{\sigma}}_{ij} &= \sigma(c_is_je^{-i\sigma\phi_j} -
c_js_ie^{-i\sigma\phi_i}), \nonumber
\end{align}
where $\bar{\sigma} = -\sigma$ and $c_i=\cos \theta_i/2$, $s_i=\sin
\theta_i/2$ and the set of angles $\{\theta_i \}$ and $\{\phi_i \}$ are the polar and azimuthal angles corresponding to $\{ {\bf S}_i \}$, respectively. As one can see, these effective hoppings can become
complex, and it has been shown that non-coplanar spin configurations
can endow the electronic bands with a nontrivial
topology.\cite{Ohgushi2000, Martin2008} Additionally, the itinerant electrons mediate an interaction
between the localized spins, which typically competes with
antiferromagnetic spin-spin interactions; on frustrated lattices, this
competition can resolve itself in non-coplanar -- and thus
topologically nontrivial -- phases.\cite{Martin2008,Akagi:2010p083711,Kumar:2010p216405,Kato_FKLM_tri_2010}

\subsection{Effective bands for a Kondo-lattice model with infinite
  Hund's rule coupling} \label{sec:Jinf}

The interplay of the orbital symmetries summarized in Sec.~\ref{sec:orbs}
with the Berry phases of Sec.~\ref{sec:berry} was discussed in
Ref.~\onlinecite{Venderbos2011} for the limit of infinite Hund's
rule coupling to classical localized spins, corresponding to the
double exchange model. In this case, one only keeps the $\uparrow$
electrons parallel to the local spin-quantization axis and electrons
effectively become spinless fermions. 
For the chiral spin pattern in Fig.~\ref{fig:chiral}, which has been
found as the ground state of triangular Kondo-lattice
models,\cite{Martin2008,Akagi:2010p083711,Kumar:2010p216405,Kato_FKLM_tri_2010}
the Berry 
phases between the four sites of the 
magnetic unit cell can be parametrized as
\begin{align}\label{eq:hopp_chiral}
&u^{\uparrow\uparrow}_{1,2} =
u^{\uparrow\uparrow}_{3,4}=\frac{1}{\sqrt{3}},\quad
u^{\uparrow\uparrow}_{1,3} = -u^{\uparrow\uparrow}_{2,4} =
\frac{1}{\sqrt{3}}\\
&u^{\uparrow\uparrow}_{2,3} =
u^{\uparrow\uparrow}_{4,1}= -u^{\uparrow\uparrow}_{3,2} =
-u^{\uparrow\uparrow}_{1,4}= \frac{i}{\sqrt{3}}. \nonumber
\end{align}
It turns out that these hoppings can in fact be written in a two-site
unit cell (containing sites 1 and 2) due to an internal symmetry of the four-site
pattern,~\cite{Martin2008} see also the effective model
Eq.~(\ref{eq:Hkin}) below.

\begin{figure}
\subfigure{\includegraphics[width=0.9\columnwidth]{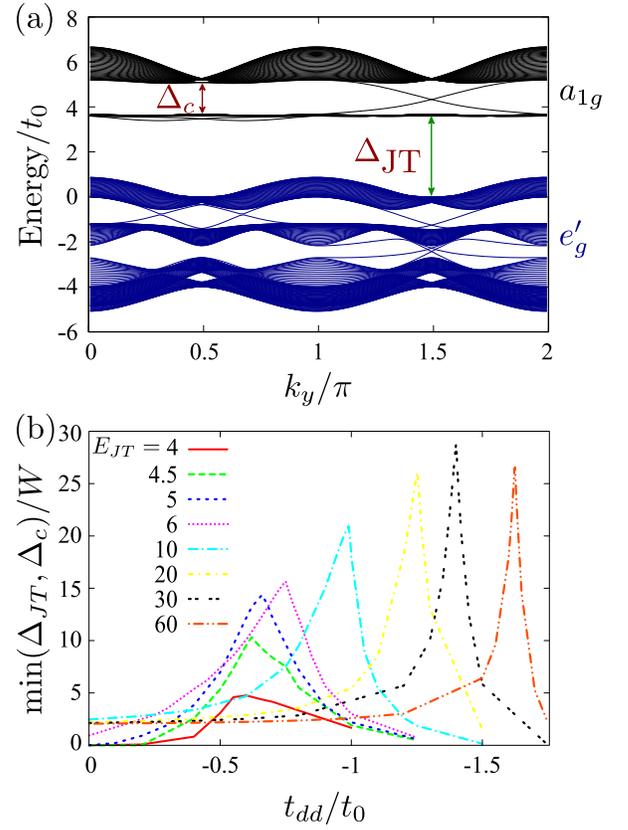}\label{fig:bands_kondo}}\\[-0.75em]
\subfigure{\label{fig:gaps_kondo}}
\caption{(Color online) Flat lower chiral subband in the Kondo-lattice
  model with infinite Hund's rule coupling (double-exchange
  model). (a) Shows the one-particle energies of three $t_{2g}$
  orbitals coupled to localized spins, where the latter form a
  spin-chiral phase on a triangular lattice,\cite{Martin2008,Akagi:2010p083711,Kumar:2010p216405,Kato_FKLM_tri_2010} see
  Fig.~\ref{fig:chiral}. The system is a cylinder, i.e., periodic
  boundary conditions along $y$-direction and open boundaries along
  $x$. The horizontal axis is the momentum in the direction with
  periodic boundaries. The gaps $\Delta_{\textrm{JT}}$ and $\Delta_c$ denote the gaps
    due to 
    crystal-field splitting $E_{\textrm{JT}}$ and to the chiral spin state. (b) shows
    the figure of merit $M$, see Eq.~(\ref{eq:merit}), for the lower
    $a_{1g}$ subband. The curves for crystal-field splittings
  $E_{\textrm{JT}}=4, 4.5$, and $5$ were already given in
  Fig.~3(b) of Ref.~\onlinecite{Venderbos2012} and are repeated here for
  convenience. 
\label{fig:kondo_inf}}
\end{figure}

Combining the phases Eq.~(\ref{eq:hopp_chiral}) with the hoppings given
by Eqs.~(\ref{eq:hopp_xyz}) or~(\ref{eq:hopp_a1g_eg}) and the
crystal-field splitting Eq.~(\ref{eq:JT_xyz}) still gives a
non-interacting model that can be easily
solved in momentum space. One finds that large
$|E_{\textrm{JT}}|$, see Eq.~(\ref{eq:JT_xyz}),  strongly reduces the dispersion of one
  subband.~\cite{Venderbos2011} This can also be seen in
Fig.~\ref{fig:bands_kondo}, which shows the one-particle energies
obtained on a cylinder. Figure~\ref{fig:bands_kondo} also reveals the
edge states crossing some gaps, indicating the topologically
nontrivial nature of these bands. Calculating Chern numbers $C$ 
corroborates this and gives $C=\pm 1$.~\cite{Venderbos2011} The band flatness can be
expressed in terms of a figure of merit
\begin{equation}\label{eq:merit}
M = \frac{\min(\Delta_{\textrm{JT}},\Delta_c)}{W},
\end{equation}
where $\Delta_{\textrm{JT}}$ and $\Delta_c$ are the two gaps
separating the narrow band of interest from the other orbitals and
from the subband with opposite Chern number and $W$ is the width of
the narrow band. As has been pointed out,~\cite{Venderbos2011} the lower
subband can here become very flat, and as can be seen in
Fig.~\ref{fig:gaps_kondo}, the flatness can be further improved by
going to larger crystal fields and reaches values $M\approx 28$. 

As these very flat bands can be achieved for large separation
$E_{\textrm{JT}}$ between the $a_{1g}$ and $e_g'$ states and as
the band of interest then has almost purely $a_{1g}$ character, it is
natural to assume that one should be able to capture the most relevant
processes with an effective $a_{1g}$ model. (This is in contrast to the
situation starting from $e_g$ orbitals, where one finds
intermediate $E_{\textrm{JT}}$ to be
optimal.~\cite{Venderbos2011} In that case, the nearly flat bands can
only be obtained if \emph{both} orbitals contribute weight and one
cannot easily reduce the situation to a one-band system.) 

The impact of the $e_g'$ levels on the effective $a_{1g}$ dispersion can  be taken into account in
second-order perturbation theory. This includes processes where a hole
hops from the $a_{1g}$ orbital at site $i$ to an $e_g'$ state at $j$
and back again to an $a_{1g}$ state at a third site $i'$, which may or
may not be the same as $i$. The denominator of these terms is the
crystal-filed energy $E_{\textrm{JT}}$ and the numerator is
obtained from the products
$\widetilde{T}_i^{ab}\widetilde{T}_j^{ba}+\widetilde{T}_i^{ac}\widetilde{T}_j^{ca}$
(with $a$ designating $a_{1g}$ and $b$, $c$ the $e_g'$ states). 
In order to evaluate the second-order hopping between sites $i$ and
$i'$, these orbital hoppings have to be multiplied by the product of
the Berry phases $u^{\uparrow\uparrow}_{i,j}$ and
$u^{\uparrow\uparrow}_{j,i'}$ from Eq.~(\ref{eq:hopp_chiral}) for all
paths connecting $i$ and $i'$ via one intermediate site $j\neq
i,i'$. Due to destructive interference, processes connecting NN and 
next-nearest neighbor (NNN) sites cancel while effective
third-neighbor hopping, where there is only one path, remains. 
Since third-neighbor spins in the chiral phase are always parallel,
the total Berry phase of this process is 1 in all directions, however,
the hopping via a spin of different orientation in the middle reduces
the hopping amplitude by $|u^{\uparrow\uparrow}|^2=1/3$, leading to 
\begin{equation}\label{eq:t3}
t_3 = -\frac{2(t_0-t_{dd})^2}{27E_{\textrm{JT}}}\;.
\end{equation}
A third-neighbor hopping $\propto \sum_i \cos 2 {\bf k}\cdot {\bf a}_i$ turns
out to have almost the same dispersion as the chiral subbands and can
consequently almost cancel it in one subband. As its strength can be
tuned by $t_{dd}$ and $E_{\textrm{JT}}$, very flat subbands can
be achieved, see Fig.~\ref{fig:gaps_kondo}.

\subsection{Impact of the upper Kondo/Hubbard band: Flat  bands for
  finite interactions} \label{sec:Jfinite}

\begin{figure}
\includegraphics[width=0.9\columnwidth]{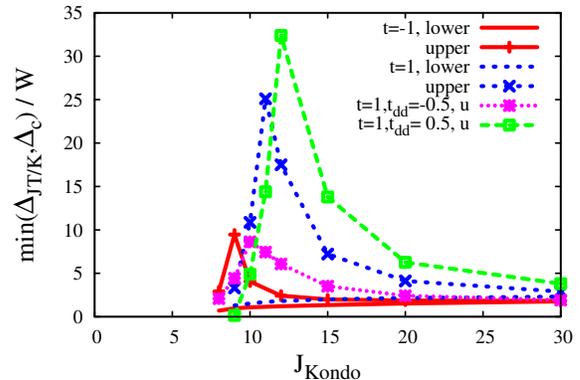}
\caption{(Color online) Figure of merit $M$, see Eq.~(\ref{eq:merit}), for
  finite Hund's-rule coupling $J_{\textrm{Kondo}}/t_0$ and
  $E_{\textrm{JT}}=6t_o$. The bands designated as ``upper'' and
  ``lower'' refer to the two subbands of the $a_{1g}$ states with
  spin parallel to the localized spin, which are separated by the gap
  opening in the spin-chiral phase, see Fig.~\ref{fig:bands_kondo}.\label{fig:gapps_kondo}}
\end{figure}

While the previous section has illustrated how one can understand
the occurrence of nearly flat bands in a three-orbital double-exchange
model, i.e., for infinite Hund's rule coupling to some 
localized spins, this section will discuss finite Hund's
rule. Figure~\ref{fig:gapps_kondo} shows the figure of merit for the
band flatness Eq.~(\ref{eq:merit}) for a few sets of hopping
parameters and for $E_{\textrm{JT}}=6t_0$ depending on Hund's rule
coupling $J_{\textrm{Kondo}}$ to the localized spin, see
Eq.~(\ref{eq:kondo_loc}).  As can be seen in 
 Fig.~\ref{fig:gapps_kondo}, the \emph{upper} subband of the $a_{1g}$
 sector can now become nearly flat. (For $J_{\textrm{Kondo}}\gg
 |E_{\textrm{JT}}|$, one can of course still find flat lower
 subbands, as discussed above.)

The flatness of the upper subband can be explained by similar
effective longer-range hoppings in second-order perturbation theory,
this time also taking into account intermediate states with an
electron in the upper Kondo band, i.e., with antiparallel spin. These
additional terms can go either via the $a_{1g}$ or via the $e_g'$
orbitals and involve combined Berry phases of the form
$u^{\uparrow\downarrow}_{ij}u^{\downarrow\uparrow}_{ji'}$. Again, one
has to sum over all possible intermediate sites $j$ and finds
\begin{align}
t_1 &= \frac{3t+\delta t}{3} + 2\frac{(3t+\delta
    t)^2}{9}\frac{1}{E_2}- 2\frac{\delta t^2}{9}\frac{1}{E_3}, \label{eq:t1full}\\
t_2 &=2\frac{(3t+\delta t)^2}{9}\frac{1}{E_2}- 2\frac{\delta
    t^2}{9}\frac{1}{E_3},\label{eq:t2full}\\
t_3 &=2\frac{(3t+\delta t)^2}{27}\frac{1}{E_2}+ 4\frac{\delta t^2}{27}\frac{1}{E_3}+ 2\frac{\delta t^2}{27}\frac{1}{E_1}\;.\label{eq:t3full}
\end{align}
The NN, NNN and third-neighbor hoppings are here denoted by $t_1$,
$t_2$, and $t_3$. $E_1=E_{\textrm{JT}}$, $E_2=J_{\textrm{Kondo}}$, and $E_3=J_{\textrm{Kondo}}-E_{\textrm{JT}}$ give the excitation energies
of the intermediate states with (i) a hole in the $e_g'$ states with
spin parallel, (ii) an electron in the $a_{1g}$ states with spin
anti-parallel and (iii) an electron in an $e_g'$ state with spin
anti-parallel. 
Like the bare NN hopping, these effective hopping-matrix elements acquire an
additional Berry phase $u^{\uparrow\uparrow}_{ii'}$ in the Hamiltonian, which only
depends on the relative orientation of spins on the initial and final
sites. NNN hopping $t_2$ via the upper Kondo band does not drop out,
and NN hopping becomes renormalized. 

The flat chiral subbands that have been observed in a three-orbital
$t_{2g}$ Hubbard model on the triangular lattice~\cite{Venderbos2012}
arise in situations similar to the finite-$J_{\textrm{Kondo}}$
scenario, and a mapping to the Kondo-lattice model was shortly
mentioned in the supplemental material of
Ref.~\onlinecite{Venderbos2012}. The key point of the mapping is the
observation that large crystal-field splitting $E_{\textrm{JT}}$, see 
(\ref{eq:JT_xyz}), leads to an orbital-selective
Mott insulator, where the $e_g'$ levels are half filled and far from
the Fermi level, while the states near the Fermi level have almost only
$a_{1g}$ character. The orbital degree of freedom is consequently
quenched, because orbital occupations are already fully
determined. A charge degree of freedom remains, as the $a_{1g}$
orbital contains one electron per two sites. Charge fluctuations of
the half-filled $e_g$ levels, however, are suppressed due to the large
Mott gap between their occupied and empty states. They can thus be
described as a spin degrees of freedom, and the situation is further
simplified, because they form a total spin with $S=1$  due to Hund's
rule. The $a_{1g}$ electron is likewise coupled via FM
Hund's rule to this spin. This situation -- mobile carriers coupled
via FM Hund's rule to localized spin degrees of freedom -- is captured
by a FM Kondo-lattice model.

These considerations can be cast in a more formal setting starting from the mean-field
decoupling~\cite{Venderbos2012} of the onsite Coulomb repulsion. The
full interaction for equivalent $t_{2g}$ orbitals reads
\begin{align}  
  H_{\rm int}& =
  U\sum_{{\bf i},\alpha} n_{{\bf i},\alpha,\uparrow}n_{{\bf i},\alpha,\downarrow} \label{eq:intra}\\
&\quad +(U'-J/2)\sum_{{\bf i},\alpha < \beta}
n_{{\bf i},\alpha} n_{{\bf i},\beta} \label{eq:inter}\\
&\quad -2J\sum_{{\bf i},\alpha < \beta}\left(
 {\bf s}_{{\bf i},\alpha} {\bf s}_{{\bf i},\beta}
 \right) \label{eq:hund}\\
&\quad +J' \sum_{{\bf i},\alpha < \beta}\left(
\cdag_{{\bf i},\alpha,\uparrow}\cdag_{{\bf i},\alpha,\downarrow}
\cnod_{{\bf i},\beta,\downarrow}
\cnod_{{\bf i},\beta,\uparrow}
      + \textrm{H. c.} \right) \label{eq:pair}\;,
\end{align}
where $U=U'+2J$ and $J'=J$ holds in the case of equivalent $t_{2g}$
orbitals. We now concentrate on the regime of interest, where the
$a_{1g}$ orbital is separated from the $e_g$ doublet by a sizable
$E_{\rm JT}$ and does not have to be equivalent. As long as
intraorbital interaction (\ref{eq:intra}) and Hund's rule
coupling (\ref{eq:hund}) dominate over interorbital interaction
(\ref{eq:inter}) and crystal-field splitting (\ref{eq:JT_xyz}), doubly
occupied orbitals will be suppressed and the last term (\ref{eq:pair})
will consequently not be important. Moreover, there is no reason for
spins in different orbitals, but on the same site, to point in
different directions, as the only interactions between spins are FM,
i.e., we can use the same local quantization axis for all
orbitals. In the case of doubly occupied orbitals, one spin can be
seen as lying antiparallel and as introduced in
Sec.~\ref{sec:berry}, ``$\uparrow$'' (``$\downarrow$'') denotes a spin
parallel (antiparallel) to the local quantization axis. The corresponding mean-field decoupling is  
\begin{align}  
  H_{\rm mf}& \approx 
  U\sum_{{\bf i},\alpha} \left(
\langle n_{{\bf i},\alpha,\uparrow}\rangle n_{{\bf i},\alpha,\downarrow}
+ n_{{\bf i},\alpha,\uparrow} \langle  n_{{\bf i},\alpha,\downarrow}\rangle\right)\label{eq:Umf}\\
&\quad +(U'-J/2)\sum_{{\bf i},\alpha < \beta}\left(
\langle n_{{\bf i},\alpha}\rangle n_{{\bf i},\beta}
 +n_{{\bf i},\alpha} \langle n_{{\bf i},\beta} \rangle \right)\label{eq:Upmf}\\
&\quad -2J\sum_{{\bf i},\alpha < \beta}\left(
\langle m_{{\bf i},\alpha} \rangle m_{{\bf i},\beta}
+m_{{\bf i},\alpha} \langle m_{{\bf
    i},\beta} \rangle  \right) + C\;,\label{eq:Hundmf}
\end{align}
where $C$ is a constant~\cite{Venderbos2012} and $m_{{\bf i},\alpha}=(n_{{\bf
    i},\alpha,\uparrow}-n_{{\bf i},\alpha,\downarrow})/2$. 
Due to the last term (\ref{eq:Hundmf}), an electron in orbital $\beta$ feels a FM
coupling to a ``classical localized spin'' with length
$\sum_{\alpha\neq \beta}\langle m_{{\bf i},\alpha}\rangle$ that 
points along the local quantization axis. The axis can be parametrized
by angles $\theta_{\bf i}$ and $\phi_{\bf i}$, which establishes the
relation to Sec.~\ref{sec:berry}. The first  
term (\ref{eq:Umf}) suppresses doubly occupied orbitals for all three
orbitals. In the orbital-selective Mott-insulator, $n_{{\bf i},\alpha}
\approx n_{{\bf i},\alpha,\uparrow}\approx 1$ for the $e_g'$ states. The second
term (\ref{eq:Upmf}) thus mainly enhances the effect of $E_{\rm JT}\to
E_{\rm JT} + 2U'-J$. The last term (\ref{eq:Hundmf}) becomes
equivalent to the Kondo term, Eq.~\eqref{eq:kondo_loc}:
\begin{align}  \label{eq:kondo_a1g}
H_{{\rm Kondo},a_{1g}}= -2J \sum_{{\bf i}}
 m_{{\bf i},a_{1g}} = -J \sum_{{\bf i}} (n_{{\bf
    i},a_{1g},\uparrow}-n_{{\bf i},a_{1g},\downarrow})\;.
\end{align}

Finally, we note that the fact that onsite interactions $U$ and
$J$ are only large, but \emph{not} infinite, is  
important for the spin-chiral ground state: for very large $U$ and
$J$, the ground state becomes a ferromagnet.~\cite{Venderbos2012} This can be related to
the fact that the Kondo-lattice model requires
either finite $J_{\textrm{Kondo}}$~\cite{Akagi:2010p083711} or
additional AFM inter-site superexchange~\cite{Kumar:2010p216405} to
support a spin-chiral instead of a FM state. At finite onsite
interactions, virtual excitations with doubly occupied $e_g'$ orbitals are
possible and lead to second-order processes that are similar to the
effective longer-range hoppings discussed above. In such a process, an
$e_g'$ electron hops into an occupied $e_g'$ state at a NN site,
creating a (virtual) intermediate state with energy $\propto
U+J\approx U'+3J$, and hops back in the next step. Such a
process yields an  energy gain $\propto t_{e_g'}^2/(U+J)$
and is only possible if the spins of the two involved electrons, which occupy the
same orbital in the intermediate state, are opposite. The
mechanism thus effectively provides the needed AFM intersite
superexchange and the spin-chiral state becomes stable for wide parameter
regimes, including ranges 
supporting nearly flat upper chiral subbands.~\cite{Venderbos2012}

\subsection{Effective Model} \label{sec:eff_model}

As we have discussed in the previous
Sec.~\ref{sec:Jfinite}, the most realistic route to nearly flat
bands with nontrivial topology on the triangular lattice arises at
finite Hubbard/Kondo coupling, where effective second-neighbor hopping
is generated in addition to NN and third-neighbor terms. However, we
have found that all essential features of the band structure can be captured by using just NN and
third-neighbor hoppings (see Figure in the supplemental material of
\onlinecite{Venderbos2012}). In the interest of simplicity, we
consequently adopt here the kinetic energy
\begin{equation}\label{eq:Hkin}
 H_{\textrm{kin}}({\bf k}) = 2t \sum\nolimits_j \sigma^j \cos {\bf k} \cdot {\bf a}_j + 2t'  \sum\nolimits_j \sigma^0 \cos 2{\bf k} \cdot {\bf a}_j\;, 
\end{equation}
where ${\bf a}_j$ ($j=1, 2, 3$) denote the unit vectors on the
triangular lattice, and $t$ and $t'$ are the NN and third-neighbor
hopping, which can be related to Eqs.~(\ref{eq:hopp_chiral}) and (\ref{eq:t3})
for the double-exchange scenario, to Eqs.~(\ref{eq:t1full}) and (\ref{eq:t3full}) for finite
onsite interactions, or which can be taken as fit
parameters. Pauli matrices $\sigma^j$ and unit matrix
$\sigma^0$ refer to the two sites of the electronic unit cell in the
chiral state.~\cite{Martin2008} The unit cell and the topologically
non-trivial bands are due to the symmetry breaking involved in the
underlying magnetic order. The dispersion of Eq.~\eqref{eq:Hkin} is
\begin{equation}\label{eq:ek}
\epsilon^\pm_{\bf k} = \pm 2t\sqrt{\sum_j \cos^2 {\bf k} \cdot {\bf
    a}_j} + 2t' \sum_j \cos 2 {\bf k} \cdot {\bf a}_j. 
\end{equation}
From now on, 
we use the effective NN hopping $t$ as unit of energy; the band
flatness can then be tuned by varying the ratio $t'/t$.

\begin{figure}[t!]
 \centering
\subfigure{\includegraphics[width=\columnwidth]{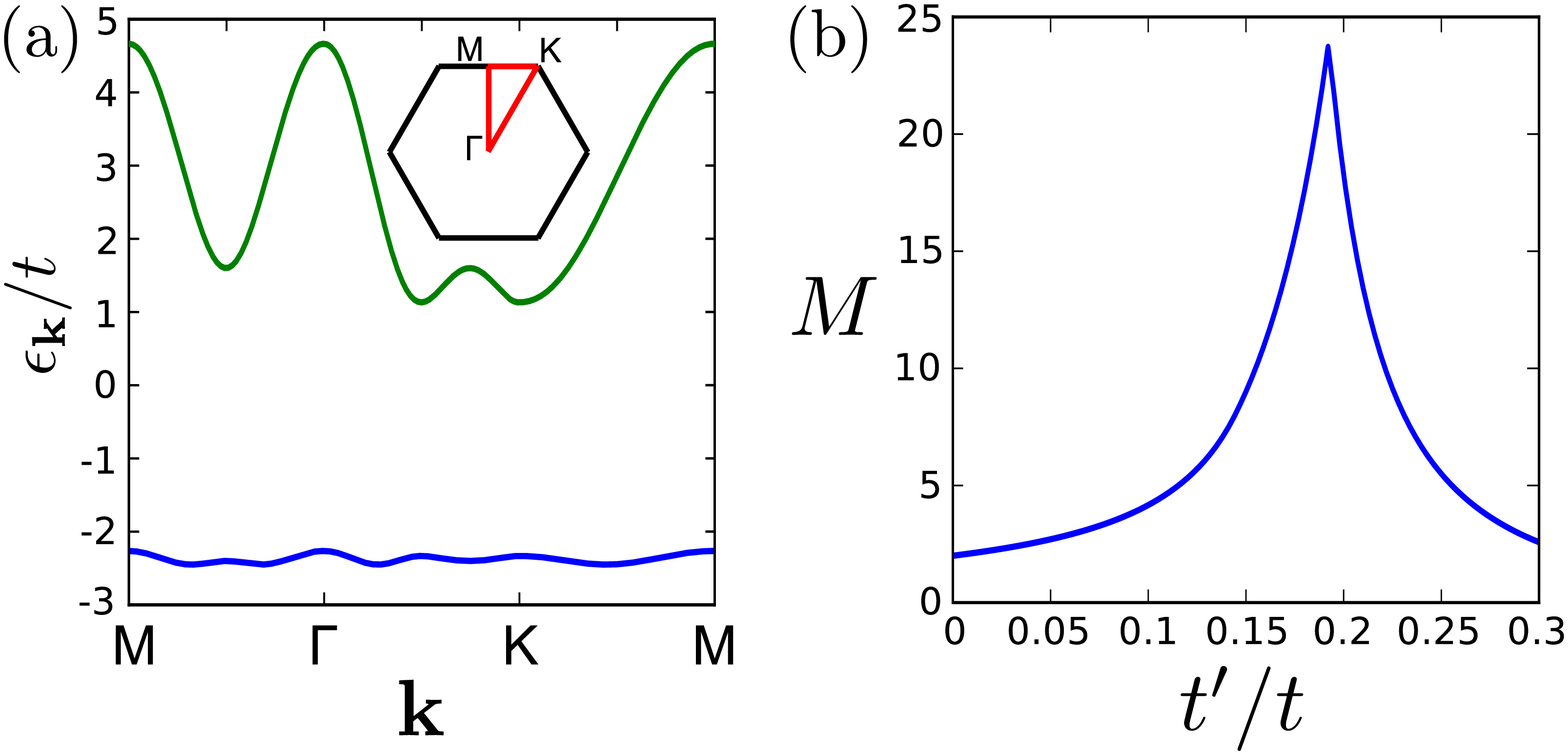}\label{fig:eff_disp}}\\
\subfigure{\label{fig:eff_flatness}}
\caption{(Color online) (a) Energy dispersion $e_{\bf k}$ for
  $t'/t=0.2$. The inset shows the path taken through the first
  Brillouin zone. (b)
  Flatness ratio $M$, see Eq.~(\ref{eq:merit}), as a function of
  $t'/t$. The flatness ratio has been calculated from the dispersion
  along the high-symmetry directions shown in the inset on the
  left.}\label{fig:disp} 
\end{figure}

The longer range hopping $t'$ determines the flatness of the bands of
$H_{\textrm{kin}}$, which can be expressed by the figure of merit $M$,
see Eq.~(\ref{eq:merit}). Figure~\ref{fig:disp} shows $M$ depending on
$t'$, and one sees that ratios $M\gtrsim 20$ can be reached for
$t'\approx 0.2$. Such flatness ratios can reasonably be achieved in
the low-energy bands of a strongly correlated $t_{2g}$ system on a
triangular lattice.~\cite{Venderbos2012} Changing the sign of $t'$ simply
mirrors the dispersion vertically, i.e., it is then the upper band
that becomes nearly flat. When going away from maximal $M$, the bands
for smaller and larger $t'$ differ qualitatively; for $t'<0.2$, the
Fermi surface (FS) at some fillings is almost perfectly nested. We are
going to discuss the impact of these differences in
Sec.~\ref{sec:phdiag}. 

The NN Coulomb interaction 
\begin{equation}\label{eq:Hint}
H_{\textrm{int}}=V\sum_{\langle i,j\rangle}n_i n_j\;,
\end{equation}
is added to the kinetic term \eqref{eq:Hkin}, giving the total
Hamiltonian $H=H_{\textrm{kin}}+H_{\textrm{int}}$. While the spin-chiral
phase providing the nontrivial topology can only be expected to remain
stable for not-too-large $V$ and doping of the flat band, we are going
to study a variety of filling and interaction ranges to obtain a
comprehensive picture of the model on finite-size systems.

\section{Results} \label{sec:results}

The interacting Hamiltonian can be diagonalized exactly for small
systems, using the Lanczos method.~\cite{Lanczos1950,Lanczos1952}
Unless otherwise noted, the results presented here are on a $4\times3$
unit-cell torus (i.e., $4\times 6$ real-space sites). We do not project onto the flat subband, see
Fig.~\ref{fig:disp}(a), but instead model the whole system of
both subbands with $C=\pm 1$. While this approach increases the Hilbert
space and thus restricts system size, it has the advantage that
competition with phases mixing the two subbands is included
automatically. We use several observables to detect FCI states and
to distinguish them from other phases; first, we discuss conclusions
to be drawn from the eigenvalues, and later also include information
obtained from the eigenstates, namely the many-body Chern number as a
topological invariant and the charge-structure factor indicating
formation of a (conventional) charge-density wave.

Even though the results of our investigation do not allow for
conclusive finite-size scaling, we have verified that the findings
presented here are consistent for smaller as well as somewhat larger
systems. Furthermore, the main features of the FCI states found are
consistent for all aspect ratios yielding the same system size, apart
from those that reduce the system to a one-dimensional chain. Before
passing, we note that we have obtained similar results to the ones
presented here for several other filling fractions $\nu=p/q$, with
$q=5,7$. Conclusive evidence for 4/5, 5/7, 6/7 states could
not be obtained, in agreement with reported results for another model.~\cite{2012arXiv1207.6094L}

\subsection{Eigenvalues and flux insertion}\label{sec:spectrum}

\begin{figure}[t!]
 \centering
\includegraphics[width=0.95\columnwidth]{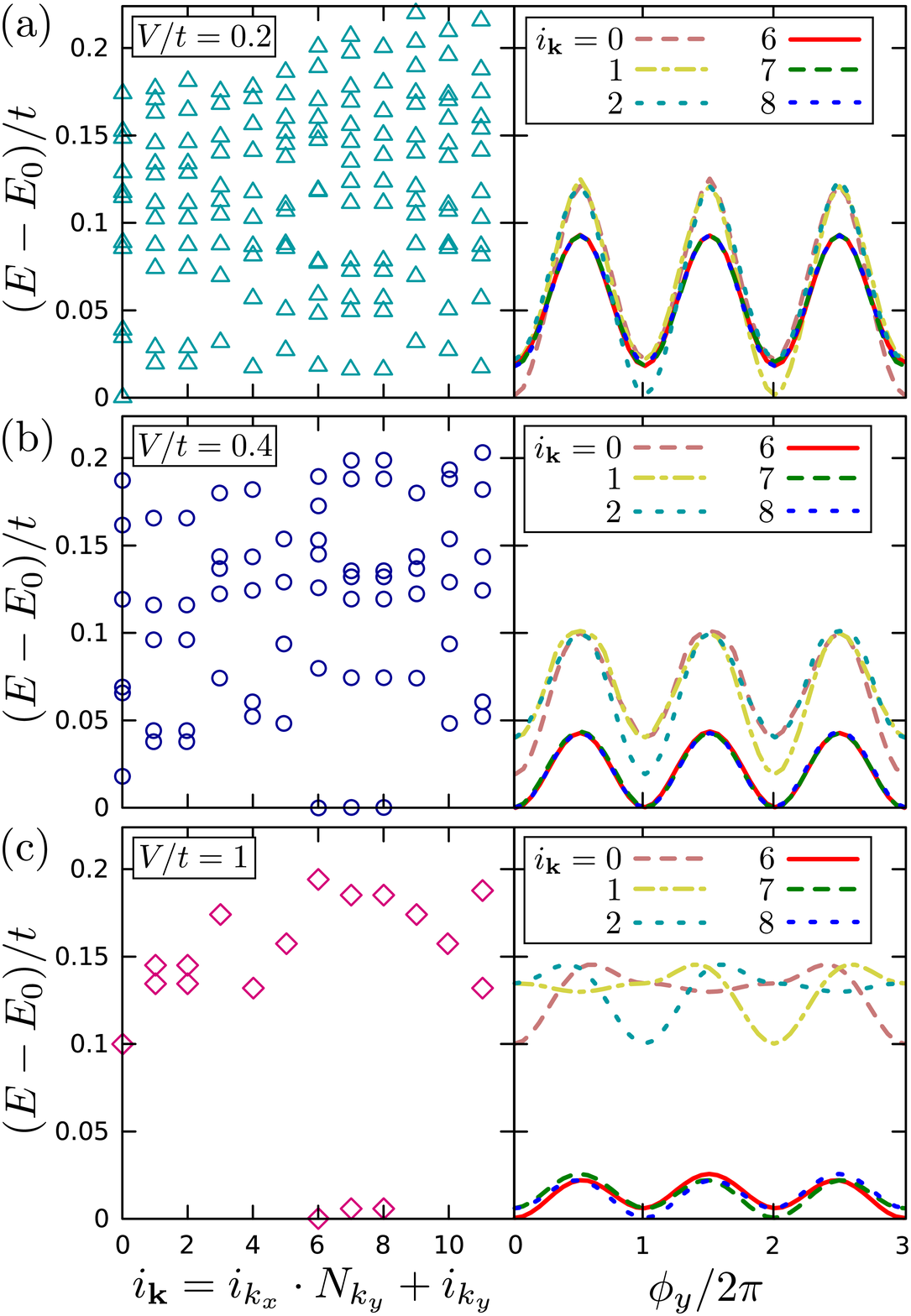}
\caption{(Color online) Eigenvalue spectrum and evolution of selected levels under flux insertion for different values of $V/t$. In all panels $t'/t=0.2$.}\label{fig:evflux_V}
\end{figure}

FCI states manifest themselves in features of the
obtained eigenvalue spectra. Traditional FQH ground states are
$q$-fold quasi-degenerate on a finite torus at filling
$\nu=p/q$.~\cite{Wen1990} The same degeneracy is expected to occur for
FCI states as well, at least within certain well-defined
limits.~\cite{Bernevig2012} Quasi-degenerate FQH ground-state
eigenvalues must also exhibit spectral flow, leading each of them into
another upon insertion of a flux quantum.~\cite{Tao1984} Magnetic
fluxes can be modelled by introducing phase factors to the hopping
from site ${\bf i}= i_x{\bf a}_1+i_y{\bf a}_2$ to site ${\bf j}=
j_x{\bf a}_1+j_y{\bf a}_2$, thus leading to the transformation
$t_{\bf{i},\bf{j}} \to t_{\bf{i},\bf{j}} \exp\left[{i\left(\phi_x
      \frac{j_x-i_x}{L_x}+ \phi_y
      \frac{j_y-i_y}{L_y}\right)}\right]$. In this manner, an electron
hopping around the lattice in the $x/y$ direction picks up a phase
$\phi_{x/y}$. 

\begin{figure}[t!]
 \centering
\includegraphics[width=0.95\columnwidth]{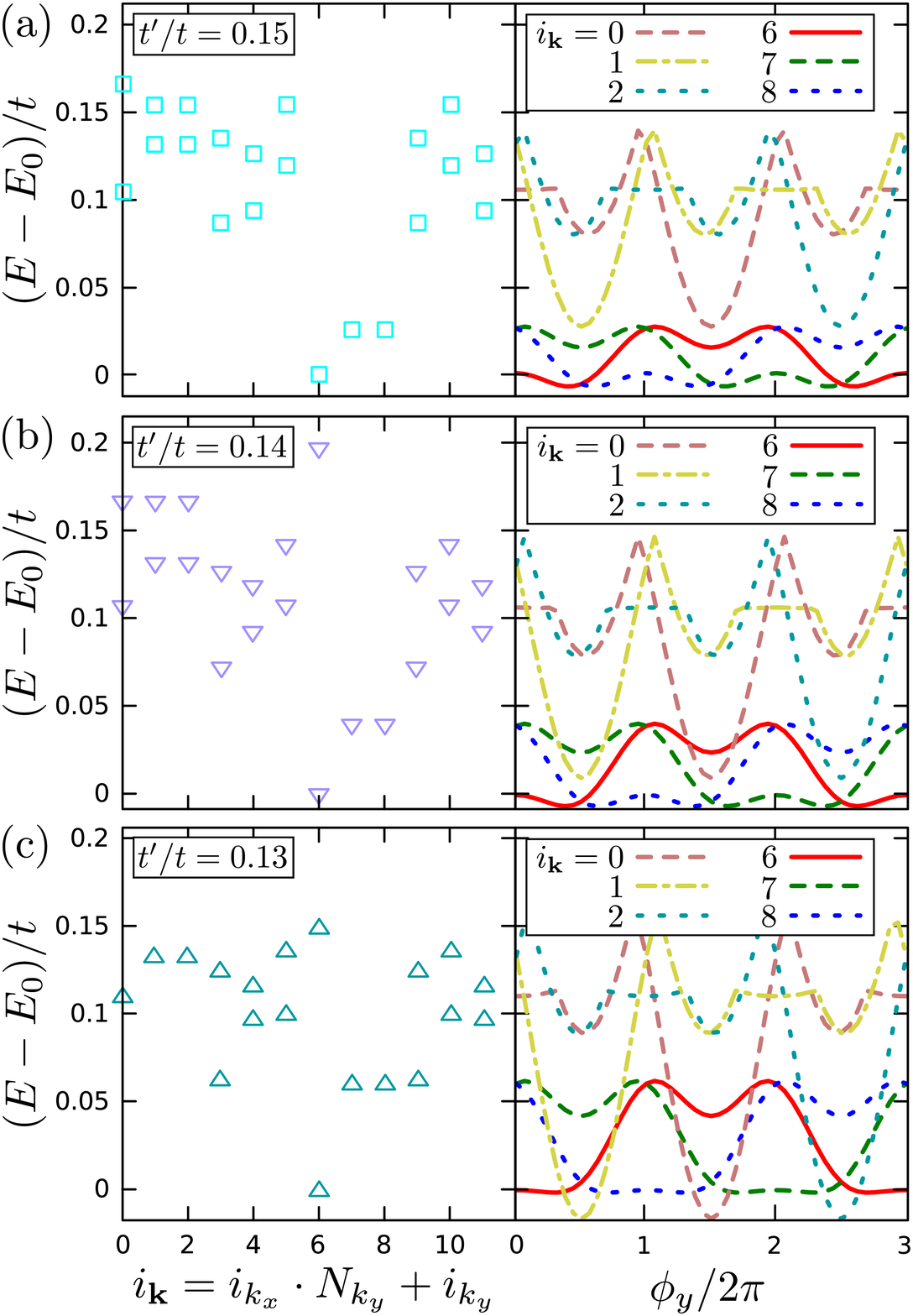}
\caption{(Color online) Eigenvalue spectrum and evolution of selected levels under flux insertion for different values of $t'/t$. In all panels $V/t=1$.}\label{fig:evflux_t3}
\end{figure}

The features described above are illustrated, as a function of $V/t$,
in Fig.~\ref{fig:evflux_V} for $\nu=1/3$. In this case, the three
quasi-degenerate ground states emerge from the continuum as the interaction
strength is increased. Insertion of one flux quantum
indeed leads from one ground-state eigenvalue to another. It can also
be seen that spectral flow is a general property of the eigenvalue
spectrum, and we have found that it also occurs even at
$V/t=0$. The $\nu=2/3$ ground states behave in a similar fashion. 
It can be seen that the FCI states remain gapped for a
range of both $M$ and $V$. In order to define a physically meaningful
phase boundary, however, one has to make sure that the ground states
remain gapped upon flux insertion, not only in their own momentum
sector, but also across momentum sectors. The reason is that the
presence of impurities would mix the momentum sectors and therefore
close the gap if the quasi-degenerate FCI-state levels cross excited-state levels.

As seen in Fig.~\ref{fig:evflux_V}, interaction mostly changes the
energy differences between groups of three quasi-degenerate
eigenstates, which flow into each other upon flux insertion, and has less impact
on the energy splitting within each manifold. When tuning the
transition via increasing the band dispersion, the situation is somewhat different, see
Fig.~\ref{fig:evflux_t3}. For $M\approx 10$, the FCI ground state is
destroyed by increasing the split between
quasi-degenerate levels, permitting a higher-energy states to mix with
the FCI-state manifold for some flux values. The splitting between ground states in the traditional
FQHE case is due to quasiparticle-quasihole excitations propagating
around the torus and therefore depends mainly on system size, becoming
zero in the thermodynamic limit.~\cite{Wen1990} In the finite
tight-binding lattice systems discussed here, the splitting is affected
significantly by the residual kinetic energy of the partially filled band, as
can be seen in Fig.~\ref{fig:evflux_t3}. Since the dispersion is present also
in the thermodynamic limit, it is not clear whether the splitting
survives to the thermodynamic limit or not in this system.

Despite level rearrangements or increasing spread between eigenvalues,
FCI states remain topologically conjugate upon varying the band
flatness or the interaction strength. A factor that can split this
conjugacy is disorder, as will be shown in the
Sec.~\ref{sec:disorder}. At the same time, ground-state levels that remain
gapped upon flux insertion along one direction may cross excited-state levels upon flux insertion along another direction. A more careful
analysis, based both on eigenenergy as well as on eigenstate
properties, is necessary to uniquely determine a gapped ground state
for a given parameter set. This is the purpose of
Sec.~\ref{sec:chern}. Before investigating eigenstates, however, the
next section is going to discuss another aspect of eigenenergies,
fractional statistics.

\subsection{Fractional statistics} \label{sec:stat}

\begin{figure}[t!]
 \centering
\includegraphics[width=\columnwidth]{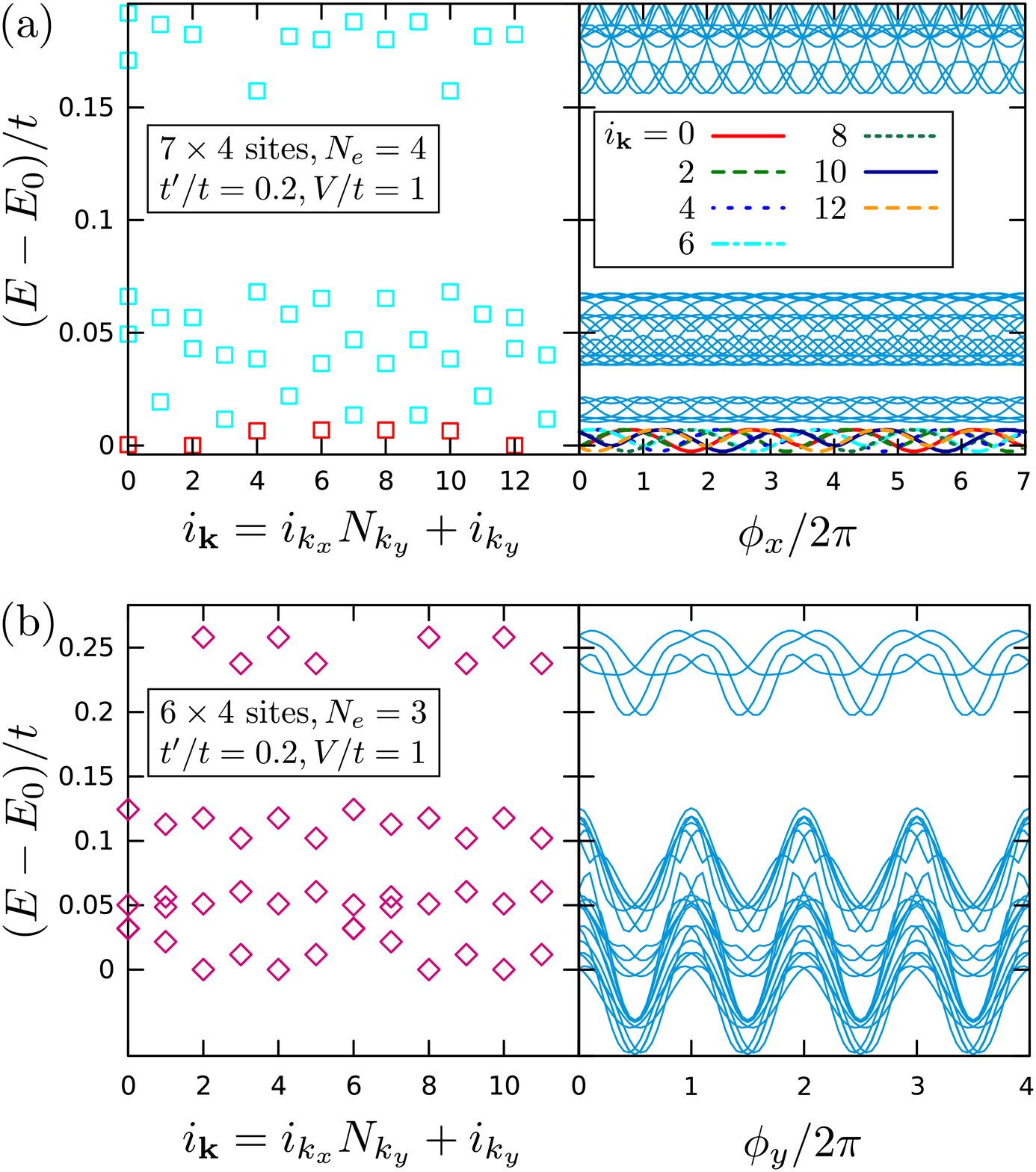}
\caption{(Color online) Eigenenergy spectrum and spectral flow for FCI state with
  quasiholes introduced (a) by enlarging the system or (b) by removing
  particles. The numbers of levels below the legends are (a) 35
  and (b) 40 respectively, in agreement with the counting
  rule of Ref.~\onlinecite{Bernevig2012}. In the $7\times2$-unit cell system,
  the filling fraction is $\nu=2/7$. There are seven quasi-degenerate
  ground states (marked in red), which are slightly separated from the
  rest of the states in the low-energy sector and exhibit spectral
  flow (shown in different colors). Their Chern number is $C=2/7$
  within the numerical error margin.}\label{fig:fractional} 
\end{figure}

The concept of generalized Pauli principles,~\cite{Haldane1991}
according to which states with clustered anyonic quasiparticles are
energetically penalized, has served as a tool to indicate fractional
statistics in FQH states.~\cite{He1992,Johnson1994} As was first
demonstrated heuristically~\cite{Regnault2011} and later on
substantiated theoretically,~\cite{Bernevig2012} the same logic holds
for FCI states. For the $\nu=1/3$ FCI state, the number of
quasihole-state levels in a well separated, low-energy Fock space is
equal to the number of (1,3)-admissible partitions of the momentum
sectors on the torus. The individual states per momentum sector can
also be accounted for, due to an emergent translational symmetry (see
Ref.~\onlinecite{Bernevig2012} for details). This symmetry is a
characteristic feature of all the FCI states we have found in the
triangular lattice model. In this section, we demonstrate that
quasihole states of the triangular-lattice model obey the same
state-counting arguments. 

Quasiholes can be introduced in a FCI state by either removing
electrons or increasing the system size. An example of both cases is
presented in Fig.~\ref{fig:fractional}. The counting rule is verified
in both cases. The counting rule can also be easily generalized to all
Laughlin filling fractions $\nu=1/q$ and we have verified it for
$\nu=1/5$ in this model. 

Apart from the agreement to the counting rule, one can also notice the
emergence of a ``daughter'' FCI state within the gapped low-energy
sector, as expected in the hierarchy picture. Even though in the
present context this is mainly a peculiarity of the small system
sizes, it can nevertheless be viewed as a simple example of the
formation of a hierarchical FCI state among the quasiparticle states
of a ``parent'' FCI state. Taking this observation one step further,
it can be seen that the spectrum looks qualitatively different under
flux insertion in the two cases presented in
Fig.~\ref{fig:fractional}, namely, the eigenvalue spectrum in the case
where the filling is $\nu=1/4$ looks more like that of the
non-interacting system with levels crossing upon flux insertion,
whereas at $\nu=2/7$ levels come in small groups, which remain
separated under flux insertion (compare to
Fig.~\ref{fig:evflux_V}). This observation can be compared to the
composite fermion theory of the FQHE,~\cite{Jain1989} according to
which the state of a FQH system at filling fractions with even
denominators can be effectively described by free fermions with flux
tubes attached to them. This composite-fermion view of the FCI hierarchy
of states has also been supported by recent numerical
calculations.~\cite{2012arXiv1207.6094L} These two different facets of the same model
can serve as an example of the compatibility between the hierarchy and
composite-fermion pictures of the FQHE in an unconventional FQH
system.

\subsection{Topological invariant} \label{sec:chern}

The most unambiguous characteristic of a FCI state is arguably its
Hall conductivity. The ground-state Hall conductivity of a FCI at filling
$\nu=p/q$ should be exactly $\nu$. Occasionally, this quantity is referred to as ``fractional Chern number'', to highlight the connection to the topological aspects of FCI states. The Hall conductivity $\sigma$ of a gapped,
degenerate state, measured in units of $e^2/h$ in the following, can be efficiently calculated using the Kubo
formula:~\cite{Niu1985,Xiao2010}  
\begin{equation}
 \sigma = \frac{N_c}{\pi q} \sum_{n=1}^q \iint_0^{2\pi} d\phi_x d\phi_y \Im \sum_{n'\not=n} \frac{ 
   \langle n | \frac{\partial { H}}{\partial \phi_y} |
     n' \rangle 
   \langle n' | \frac{ \partial { H}}{ \partial
     \phi_x} | n\rangle}
{(\epsilon_n -
   \epsilon_{n'})^2},\label{eq:chern}
\end{equation}
where $N_c$ is the number of unit cells, $\ket{n}$ are the $q$-fold
degenerate many-body ground states and $\ket{n'}$ are higher-energy
eigenstates. The corresponding eigenenergies are
$\epsilon_{n/n'}$. The $\phi_{x/y}$ dependence of the Hamiltonian
comes from magnetic fluxes going through each of the handles of the
torus, as discussed in Sec.~\ref{sec:eff_model}. 

\begin{figure}[t!]
 \includegraphics[width=\columnwidth]{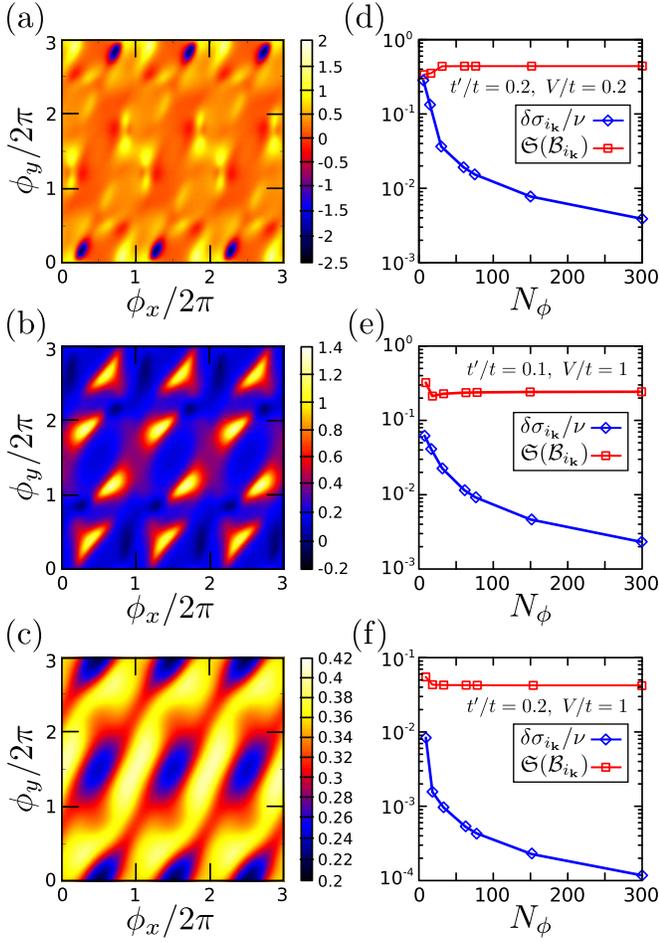}
\caption{(Color online) (a-c) Berry curvatures ${\cal B}_{i_{\bf k}}$, (d-f) their standard deviations $\mathfrak{S}({\cal B}_{i_{\bf k}})$ from the exact
  average and relative deviations of Hall conductivities $\delta\sigma_{i_{\bf k}}/\nu$ as a function of
  grid size for the state $i_{\bf k}=6$, when it is in the excited-state
  quasi-continuum (top) and in the ground-state manifold (middle and bottom). The
  integrations in Eq.~\eqref{eq:chern} have been approximated by
  simple Riemann sums. $N_\phi$ is the number of $\phi_{x/y}$ points
  taken in the range $[0,6\pi)$ in each direction.}\label{fig:berry} 
\end{figure}

The integrand in Eq.~\eqref{eq:chern} is proportional to the Berry
curvature ${\cal B}_n$ for each of the states in the degenerate ground
state. For FCI states, Berry curvatures are periodic functions varying
with $\phi_{x/y}$. 
In Fig.~\ref{fig:berry}, the many-body Berry curvature for a specific
state, which develops into one of the FCI ground states, is shown. It
is seen that its smoothness varies as this state emerges from the
excited state continuum. The period extends over
$q$ flux quanta in one direction and one flux quantum in the other,
remains unchanged for all values of $V$ and it is the same for all
states in the ground-state manifold. The Berry curvature of the other
two ground states is the same function, but translated by one and two
flux quanta in the $\phi_y$ direction respectively. Even when the many-body
Berry curvature of a FCI state is strongly varying, it is often
centered around the value corresponding to the filling fraction and,
since it is periodic, the Hall conductivity obtained by integration is
very close to the quantized value. This occurs even if the FCI state is not
a ground state. In the example presented in Fig.~\ref{fig:berry}, the
Hall conductivity is in all three cases equal to $\nu=1/3$ within the numerical
accuracy, which we will now discuss. 

\begin{figure}[t!]
 \centering
\includegraphics[width=\columnwidth]{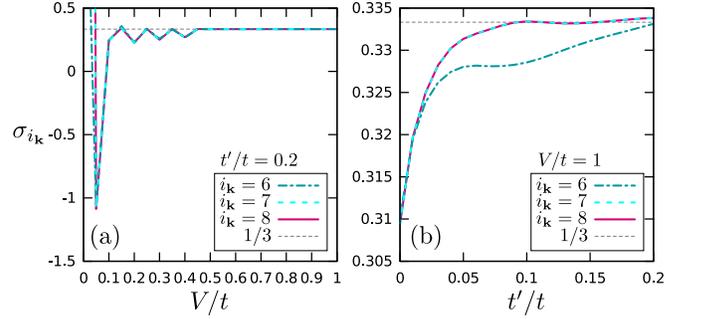}
\caption{(Color online) Hall conductivity, see Eq.~(\ref{eq:chern}), for each of the three quasi-degenerate FCI states as a
  function of (a) the interaction strength $V/t$ and (b) the
  third-neighbor hopping $t'/t$ (right). Not the different scales on
  the vertical axes.}\label{fig:chern} 
\end{figure}

The Hall conductivities we have calculated are very close to the expected
exact values in the FCI regime, despite the fact that the integrations
involved are performed numerically. The relative deviation with
respect to the exact value for the case of $\nu=1/3$ is presented in
Fig.~\ref{fig:berry}. This example illustrates that the error in the
Hall conductivity due to the finite size of the system should be smaller
than 1\%, at least for the Kubo formula approach, even for small
systems. In the case of clearly gapped FCI ground states, the accuracy
is even better, due to the integrand being very smooth. In all cases,
we have used simple Riemann summation to evaluate integrals, in order
to obtain upper bounds for the numerical errors. Other methods, like
the Simpson rule, would converge for smaller grid sizes. It should be
noted that when the Hall conductivity of an individual state within the
degenerate ground state is evaluated, the integration should be
extended over the whole period of the Berry curvature. This is no
longer necessary when calculating the average Hall conductivity and the
integration range can then be restricted to $[0,2\pi)^2$. 

The effects of interaction and band dispersion can also be traced in
the behavior of the Hall conductivity. This is shown in
Fig.~\ref{fig:chern}. Outside the FCI regime, where low-energy levels
cross upon flux insertion, the Hall conductivity oscillates. It converges
to $\nu$ at about the value of $V/t$ for which the FCI states separate
from the excited-state spectrum, forming a three-fold degenerate
ground state. On the other hand, the Hall conductivity of these states
remains close to 1/3 with deviation from the flat-band limit, and only
changes smoothly with $t'/t$, converging to the expected value close
to the phase boundary discussed in Sec.~\ref{sec:phdiag}.

\subsection{Disorder} \label{sec:disorder}

\begin{figure}[t!]
 \centering
\includegraphics[width=\columnwidth]{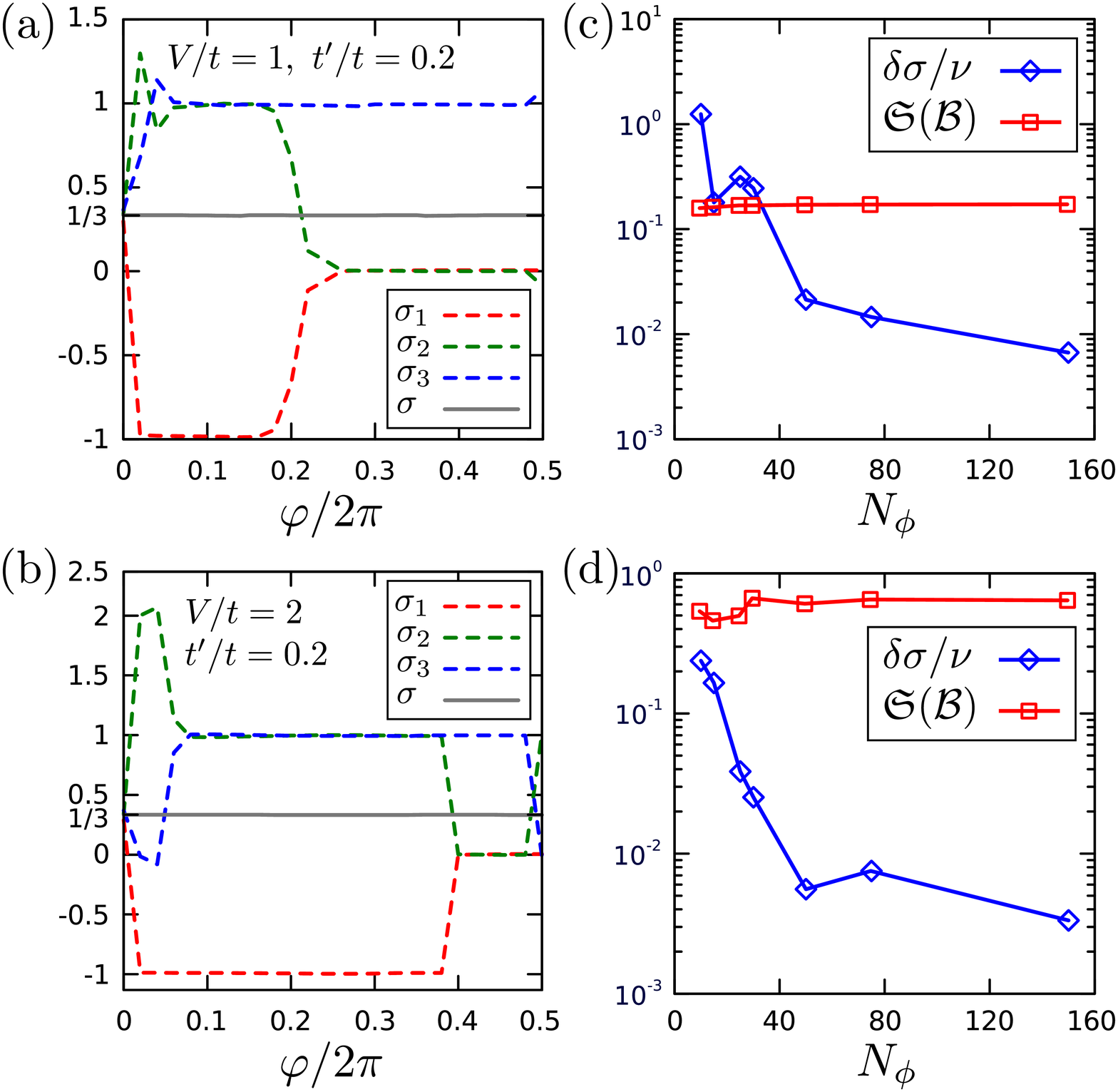}
\caption{(Color online) (a,b) Hall conductivities of the three lowest-energy states and their average as a function of disorder phase $\varphi$. (c,d) Standard deviation $\mathfrak{S}({\cal B})$ of total Berry curvature ${\cal B}$, defined by adding up the many-body Berry curvatures ${\cal B}_n$ of the three lowest-energy states for each value of $\phi_{x/y}$, and relative deviation $\delta\sigma/\nu$ of Hall conductivity as a function of grid size. $N_\phi$ is the number of $\phi_{x/y}$ points taken in the range $[0,2\pi)$ in each direction. The results shown in the right panels are for $\varphi \simeq 2\pi/3$, but are qualitatively the same for all other values of $\varphi$.}\label{fig:disorder}
\end{figure}

FQH states must be robust against disorder, as long as the energy scale of the disorder is smaller than the gap.~\cite{Wen1990} The magnetic texture in the three-orbital model presented is generated spontaneously by itinerant electrons, so an expected source of disorder are local inhomogeneities in the chiral spin pattern. A simplistic approach to simulate this effect is to vary the flux picked up by an electron hopping along one selected bond of the finite cluster. This is done by varying the phase $\varphi$ in the phase factor in front of the corresponding hopping. The effect of such a variation on Hall conductivities at $\nu=1/3$ is shown in Fig.~\ref{fig:disorder}.

As long as the energy scale of the disorder is small enough, the ground states remain separated by a gap from the rest of the spectrum and their average Hall conductivity remains constant. The Hall conductivities of the individual quasi-degenerate ground states are not smooth functions of magnetic disorder. In particular, as soon as the impurity is switched on, the Hall conductivity of each individual state jumps to an integer value. Also, Berry curvatures are no longer smooth functions of $\phi_{x/y}$. Nevertheless, the Hall conductivities are still numerically well defined, as demonstrated in the right panels of Fig.~\ref{fig:disorder}, and their average, which is the proper observable quantity in the thermodynamic limit, remains constant as the disorder is varied. This invariance of the Hall conductivity directly demonstrates the topological robustness of FCI states.

Again in analogy to FQH states,\cite{Sheng2003} disorder introduces further splitting, apart from the one due to dispersion discussed in Sec.~\ref{sec:spectrum}, between FCI ground-state eigenvalues, which no longer exchange places under flux insertion. This splitting is expected to be a finite-size effect and should disappear in the thermodynamic limit. A conclusive proof of this statement is however beyond the scope of this work. Despite the similarities, the impact of disorder in general on the energy-scale balance necessary for FCI states is qualitatively different than the situation in traditional FQH states. For example, disorder may have a different effect depending on the interaction strength and range, and may lead to another, possibly topologically trivial, state, as has been shown in a recent study on the checkerboard lattice involving chemical-potential disorder.~\cite{Yang2012}

\subsection{Competition with Charge-density wave and Phase diagram} \label{sec:phdiag}

Early on, it was pointed out that an important
prerequisite for the emergence of a FCI state is the balance
between three energy scales, namely the width of the
topologically non-trivial band, the gap(s) separating it from other bands
and the Coulomb interaction strength.~\cite{Tang2011,Sun2011,Neupert2011,Sheng2011} If the ratio between the first
two, as expressed by the figure of merit Eq.~(\ref{eq:merit}), is
large, then the interaction can become strong enough to induce FCI
states, but remain small compared to the band gaps, which avoids
mixing 
in wave function of different topological character. Many investigations have been focused on
models in the perfectly flat band limit (exactly as in a Landau level), supplemented by various types
of interaction. Rather recently, it was reported that the band flatness
alone is not in fact a reliable indicator for the stability of FCI 
states.~\cite{2012arXiv1207.4097G}

\begin{figure}
\centering
\subfigure{\includegraphics[width=0.48\columnwidth]{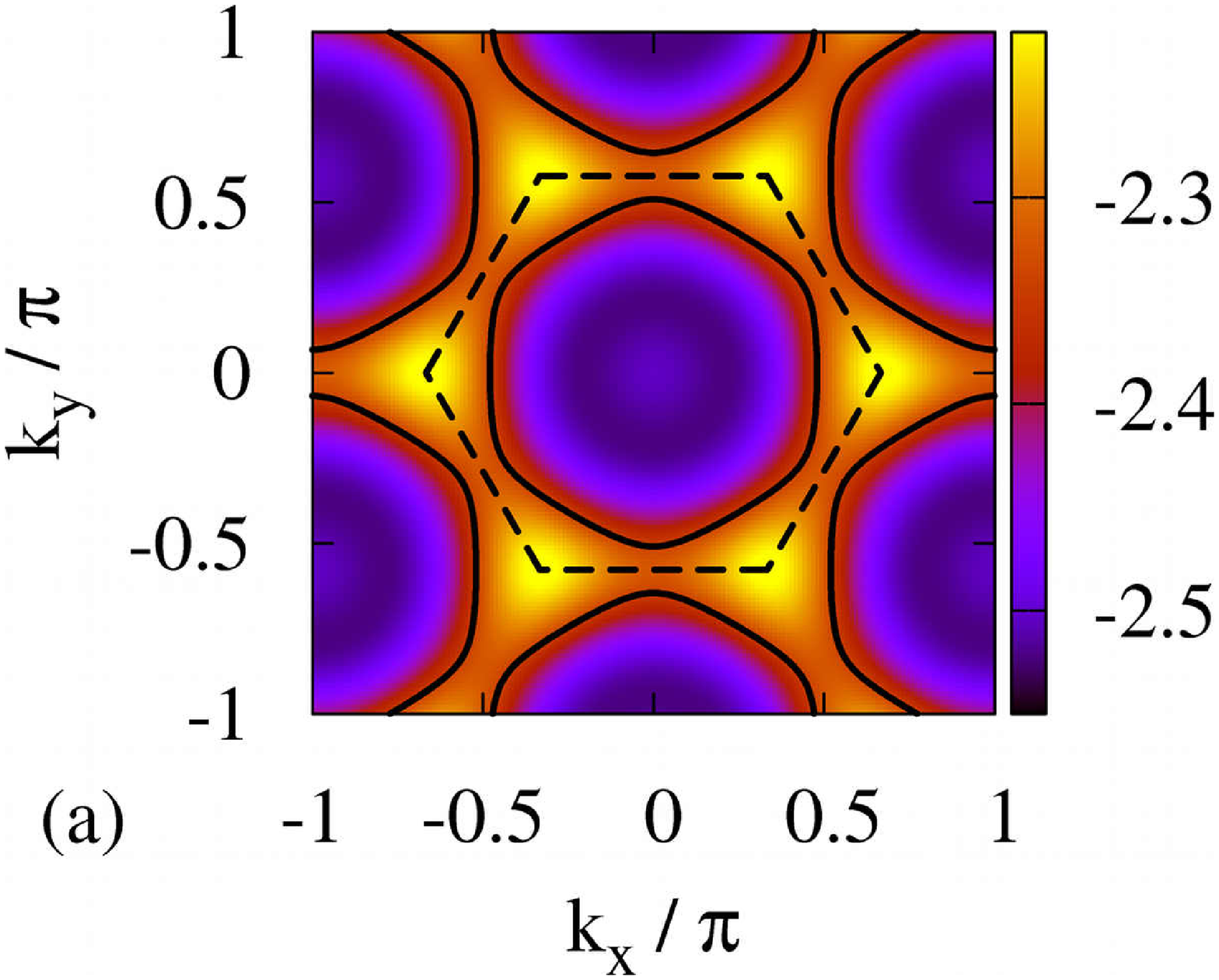}\label{fig:FBZ_FS_016}}\hfill
\subfigure{\includegraphics[width=0.49\columnwidth]{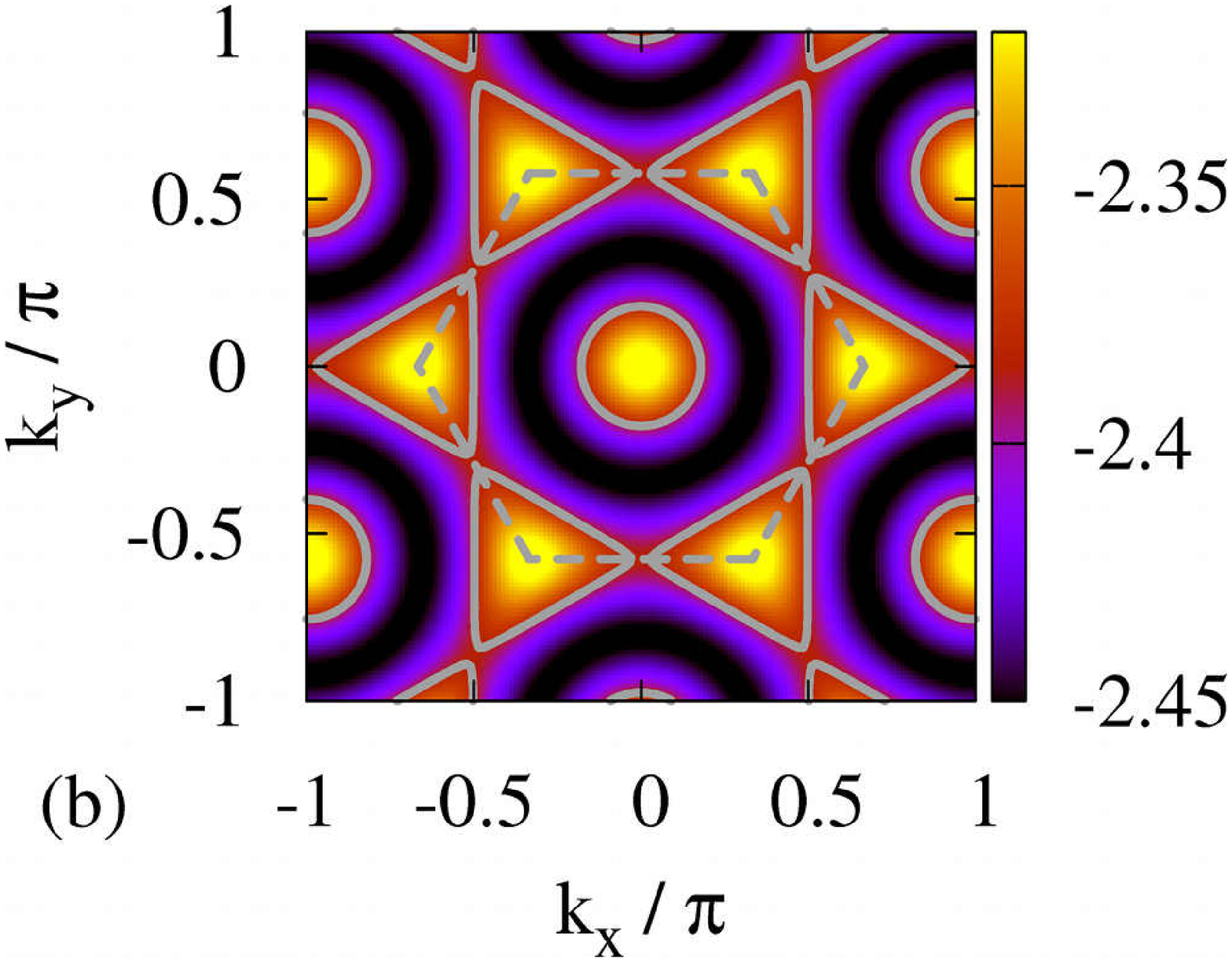}\label{fig:FBZ_FS_opt}}\\
\subfigure{\includegraphics[width=0.48\columnwidth]{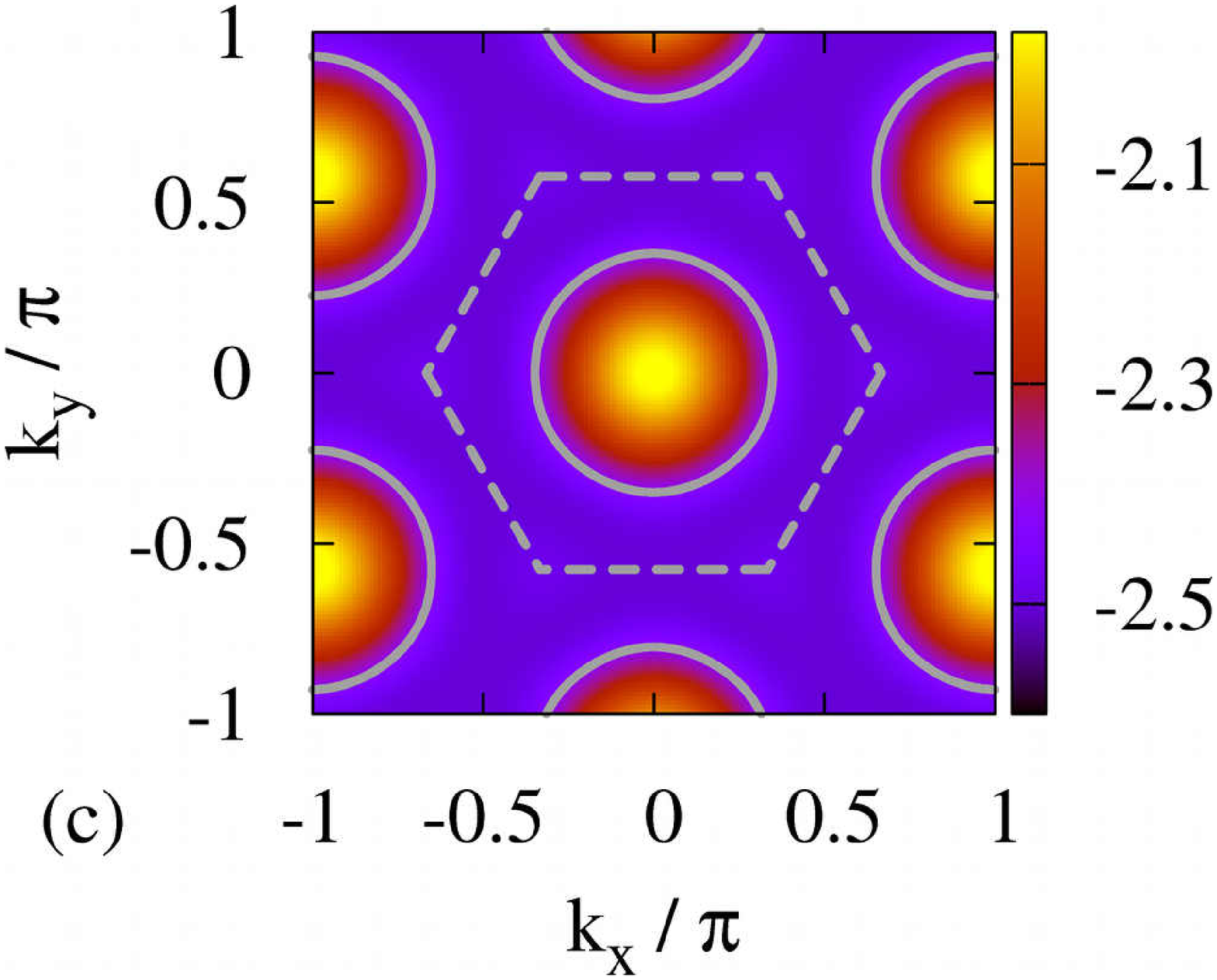}\label{fig:FBZ_FS_025}}\hfill
\begin{minipage}[b]{0.4\columnwidth}
\subfigure[]{\includegraphics[width=\textwidth]{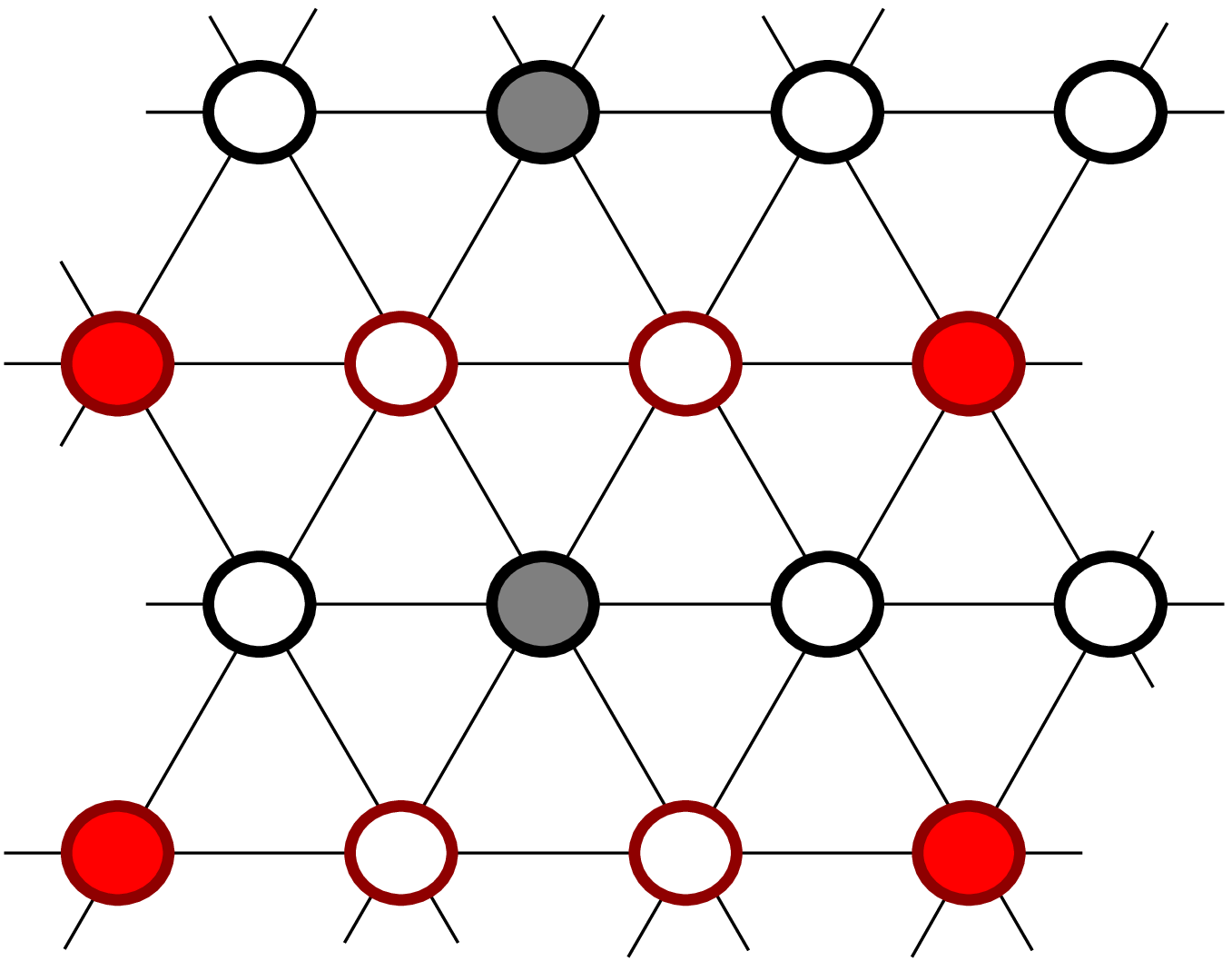}\label{fig:tri_CDW}}\\[2em]
\end{minipage}\\
\caption{(color online) Dispersion (gray-/ color-scale) and Fermi
  surface (FS) for $\nu=2\bar{n}=2/3$
  (thick solid line) for (a) $t'/t=0.16$ and $M=11.16$, (b)
  $t'/t=0.19245$ and $M=24$ (very close
  to maximal $M$) and (c) $t'/t = 0.25$ and $M=5.5$. Dashed
  lines indicate the first Brillouin zone. (d) illustrates the CDW
  possible for $\nu=2\bar{n}=2/3$. Black  and red/gray circles
  indicate the two sublattices of the spin-chiral order and the
  effective model Eq.~(\ref{eq:Hkin}), filled circles the particles
  in the CDW state induced by large $V/t$. \label{fig:FBZ_FS}} 
\end{figure}

In this section, we discuss how a finite dispersion influences the
stability of FCI states in a triangular-lattice model, where we focus
on band filling and FS nesting. The first presents a rather
obvious difference to LLs, as has recently also been
mentioned in Ref.~\onlinecite{2012arXiv1207.6094L}: while a LL is expected to be particle-hole
symmetric, the nearly flat subband of a lattice model is not. We are
going to discuss filling fractions $\nu =1/3$ and $\nu=2/3$ that turn
out to clearly exemplify this difference. The
latter case, which was also shortly discussed in
Ref.~\onlinecite{Venderbos2012}, corresponds to $1/3$ filling of the
original triangular lattice, where NN Coulomb interaction can
stabilize a CDW [with particles at second-neighbor sites, see Fig.~\ref{fig:tri_CDW}], while no
such CDW is possible at $\nu=1/3$. The competition between the FCI and
the CDW is in turn strongly affected by FS nesting, and we
are going to see that the FCI is far more stable for more dispersive bands
without nesting than for flatter bands with a nested Fermi
surface. However, we are also going to see that very flat bands allow
for FCI states at the lowest interaction strengths, even though they
are in our case very well nested.

\begin{figure}[t]
 \includegraphics[width=.8\columnwidth]{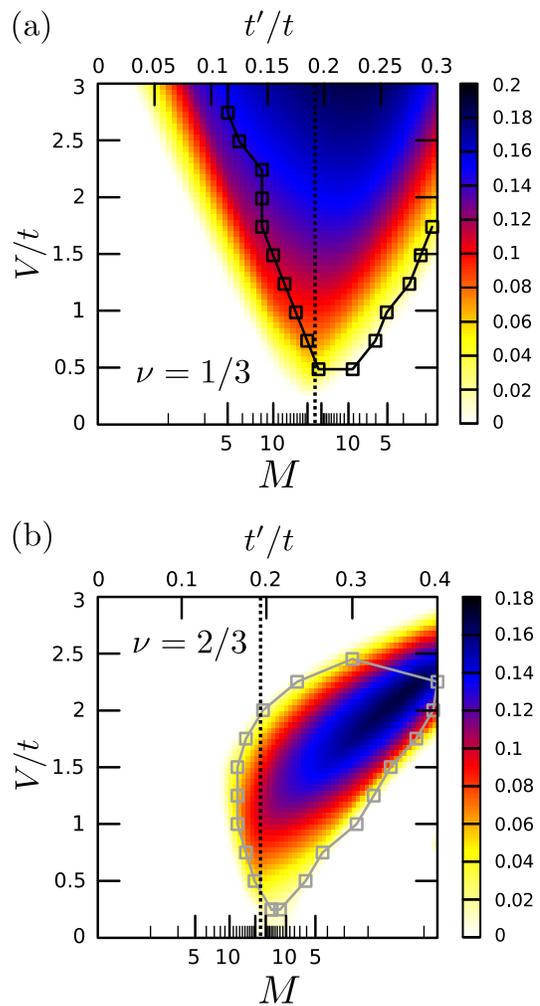}
\caption{(Color online) Gap of three-fold degenerate FCI ground state as a function
  of $M$ and $V/t$, for (a) $\nu=1/3$ and (b) $\nu=2/3$, shown
  in colorcode. The dashed lines indicate the phase boundary, defined
  by the condition that the gap remains open for all values of
  inserted magnetic flux. The flatness ratio $M$ of the flat band of
  $H_{\textrm{kin}}$ (bottom scale) is adjusted by varying $t'/t$ (top
  scale). The maximum value of the flatness ratio ($M\simeq24$) is
  marked by the dotted lines.}\label{fig:phdiag} 
\end{figure}

As mentioned in Sec.~\ref{sec:eff_model}, varying $t'/t$ allows us to
tune the band flatness and to moreover switch between regimes with
and without FS nesting. Examples are shown in
Fig.~\ref{fig:FBZ_FS}, where the dispersion as well as the Fermi
surface corresponding to $\nu=2/3$ are shown for some values of
$t'/t$. Both for near-optimal $t'/t= 0.19245$ ($M=24$) and for smaller
$t'/t=0.16$ ($M\approx 11$), the FS contains hexagons with
almost perfectly nested segments. A difference between the two cases
is on one hand the flatness ratio, but on the other hand, the flatter
bands also have an additional circular FS around the $\Gamma$
point. For $t'/t\gtrsim 0.23$, only the circular FS remains (see the
example with $t'/t=0.25$ and $M=5.5$) and there is thus no longer good
nesting. 

Figure~\ref{fig:phdiag} shows the region where FCI states are stable
on a 24-site ($4\times3$ unit-cell) system, as a function  of
Coulomb interaction strength $V$ and the flatness ratio $M$, which is
in turn controlled by third-neighbor hopping, see Fig.~\ref{fig:disp}.
The two panels are for fillings $\nu=1/3$ and $\nu=2/3$. In both
cases, the ground-state manifold in the FCI state is expected to be
three-fold degenerate. The colorcode indicates the gap between the
three ground states and the fourth lowest eigenstate in the
absence of applied magnetic flux. In order to define a physically meaningful
phase boundary, however, one has to make sure that the ground states
remain gapped upon flux insertion, not only in their own momentum
sector, but also across momentum sectors, see
Sec.~\ref{sec:spectrum}. The solid lines in Fig.~\ref{fig:phdiag}
indicate the phase boundary determined by taking these considerations into
account. We have verified that the phase diagrams obtained from a
30-site torus are qualitatively the same as the ones shown here. The
stability range of the FCI differs for $\nu=2/3$ and
system sizes commensurate with the CDW ($3\times 6$ and $6\times 6$
sites), but the corresponding phase-diagram area is still finite. 

\begin{figure}[t]
 \includegraphics[width=\columnwidth]{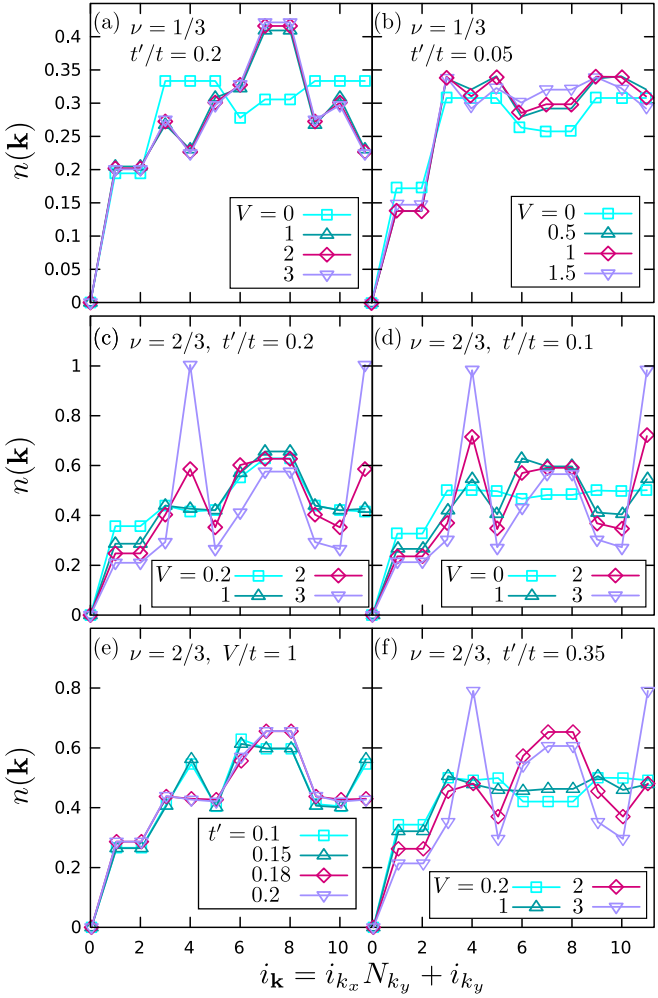}
\caption{(Color online) Static structure factor for various values of $t'/t$ and $V/t$ at
  filling fractions (a,b) $\nu=1/3$ and (c-f) $\nu=2/3$. The sharp peaks indicate the formation of a charge-ordered state. In the FCI regime, the static structure factor of all three quasi-degenerate ground states is identical.\label{fig:sf}}
\end{figure}

For both fillings, the system is in a metallic state at small $V/t$, 
while finite values of $V/t$ can lead to FCI
ground states for an extended range of the band flatness. 
In the case of $\nu=1/3$, the
FCI state can be induced for any band dispersion by making $V$ large
enough. The FCI persists even for $V/t$ considerably 
larger than the band gap. This has also been observed for another FCI model on the
checkerboard lattice,~\cite{Sheng2011} and appears to indicate that
the fermions are dilute enough to occupy mostly one subband regardless of $V/t$, so that a mixing of
the topological character of the two bands does not occur. 
In contrast, the $\nu=2/3$ FCI states survive only for moderate values of
$V/t$. Larger interaction strengths lead to a different ground state, which is, as
we will argue below, a CDW.

To determine the type of order in each of the regions in
Fig.~\ref{fig:phdiag}, we calculate key properties of the
corresponding ground states. The FCI states, despite the hints from
the energy spectrum, have to be identified by their topological
invariant, see Sec.~\ref{sec:chern}. To find out whether
there is tendency towards charge order in the rest of the phase
diagram, we calculate the static (charge-)structure factor (SSF), defined as: 
\begin{equation}
 n({\bf k}) = \frac{1}{N_s} \braket{ 0 | \sum_{j,l}^{N_s} e^{i {\bf k}
     \cdot ( {\bf R}_j - {\bf R}_l ) } (\hat n_j - \bar{n}) (\hat n_l -
   \bar{n}) | 0 }, 
\end{equation}
where $N_s$ is the number of sites, $\hat n_j$ is the electron-number
operator acting on the site at position ${\bf R}_j$, $\bar{n}=\nu/2$ is the
average electron number and $\ket{0}$ stands for the many-body ground
state. Charge-density modulations are marked by sharp features in
$n({\bf k})$ at certain wave vectors. Liquid states, on the other hand,
should be featureless in comparison to charge-modulated states. 

Even though the accessible system sizes are not large enough to
exemplify the featurelessness of liquid states, a qualitative difference between
liquid and charge-modulated states can be seen. Examples are shown
in Fig.~\ref{fig:sf}. At $\nu=1/3$ and within the FCI regime, $n({\bf
  k})$ remains almost unchanged upon variation of model
parameters. The same holds, although less markedly, for the metallic,
Fermi-liquid-like state at small interaction strengths and small (or
very large) $t'/t$. Despite the fact that the shape of $n({\bf k})$ is
distinct for the two liquid states, see Figs.~\ref{fig:sf}(a) and (b),
the differences are subtle. In order to distinguish
between the FCI and metallic states, one would thus rather use the
topological invariant of their ground states, see
Sec.~\ref{sec:chern}, or the criterion that the quasi-degenerate
ground states remain separated from higher states for all fluxes, as
used for Fig.~\ref{fig:phdiag}, see above.

However, the charge-structure factor is very valuable in distinguishing
the weakly correlated Fermi liquid from a CDW driven by Coulomb
repulsion. At $\nu=2/3$ and $t'/t=0.1<0.18$, away from the FCI regime, two 
peaks appear in $n({\bf k})$ and increase continuously upon increasing
$V/t$, see see Fig.~\ref{fig:sf}(d). 
When choosing $t'/t=0.2$ near the maximal flatness
[see Fig.~\ref{fig:sf}(c)] or $t'/t=0.35$ with bad FS nesting
[see Fig.~\ref{fig:sf}(f)], the peaks only begin to grow for large
$V$, where ground-state behavior upon flux insertion 
changes: a full gap is only obtained for lattice sizes commensurate
with the ordering pattern, like the $3\times6$ and $6\times6$
lattices, but not on more general lattices. This indicates that the
FCI state breaks down and is replaced by a CDW. The peaks observed in $n({\bf k})$ for $\nu=2/3$ grow with
$V$, supporting their relation to a CDW. Their wave vectors correspond
to a state where particles sit at NNN sites
on the triangular lattice, i.e., to the regular charge pattern
compatible with a filling of $\bar{n}=\nu/2=1/3$ of the triangular
lattice. As can be seen in Fig.~\ref{fig:tri_CDW}, this pattern avoids
any penalty due to NN Coulomb interaction $V$ and large enough $V\gg t,t'$
will thus eventually induce such a charge distribution. 

The phase diagrams in Fig.~\ref{fig:phdiag} clearly show that the band
flatness parametrized by $M$ is itself not a reliable indicator for
the stability of FCI states, as has also been pointed out recently in
a different case.~\cite{2012arXiv1207.4097G} The energy gap for
$\nu=1/3$ is rather symmetric with respect to the highest figure of
merit, but the phase boundary determined by requiring a full gap over all momentum
sectors and for all fluxes shows that the FCI states are somewhat more
stable for larger $t'/t$. 
%
In the case of $\nu=2/3$, the
asymmetry is far more striking; as one can see in
Fig.~\ref{fig:phdiag}(b), FCI states require rather flat bands with
$M\approx 13$ for $t'/t< 0.2$, but extend to a band with a
width comparable to the gap separating it from its counterpart for
$t'/t>0.2$. Having rather flat bands indeed makes it easier for small
$V/t$ to induce FCI states both at $\nu=1/3$ and $\nu=2/3$, even though
the optimal $M$ is still not quite the largest, at least for our system
sizes. As soon as the bands acquire some dispersion, however, features
beyond band flatness, in our case FS nesting, can strongly influence
the stability of FCI states by favoring competing states, in our case
a CDW. Another key feature, which extends previous
results,~\cite{2012arXiv1207.4097G} is that in the case of competition
between FCI and other phases, perfect band flatness is not necessarily
the ideal condition for the stability of FCI ground states.

Before passing, a few more comments on the phase diagram have to be
made. Ideally, a ground-state property would be used to determine the
phase boundaries. One such property, which is sensitive to phase
transitions, is the ground-state fidelity, defined as a measure of the
overlap $\braket{ \psi(\alpha) | \psi(\alpha+\delta\alpha) }$, where
$\psi$ is the ground-state wave function and $\alpha$ is a control
parameter varied in small steps $\delta\alpha$. Phase transitions are
then marked by a divergence in the fidelity at the transition
point. Having calculated the ground-state fidelity upon varying
interaction strength and band flatness for the cases presented here,
we find that such divergences occur only at points where ground-state
level crossings also occur, so the fidelity does not provide any extra
information compared to the eigenvalue spectra. 
Furthermore, it has recently been shown that the fidelity on finite
systems can fail to register topological phase transitions, e.g.,
between a FQH-like state and a Fermi liquid.~\cite{Yang2012} 

\section{Summary and Conclusions} \label{sec:conclusions}

Based on an earlier investigation,~\cite{Venderbos2012} we have illustrated extensively how FCI states can emerge in a 
strongly correlated multiorbital model and have
shown how an effective spinless one-orbital model with nearly flat
bands and non-zero Chern number $C=\pm1$ arises as the the low-energy
limit of Kondo-lattice and Hubbard models for $t_{2g}$ orbitals on a
triangular lattice. NN Coulomb interaction $V$ then
stabilizes states with all the characteristics of FCI states: the
lowest eigenvalues have a near degeneracy
corresponding to the denominator of the filling fraction and show spectral
flow without closing of the energy gap. We have, moreover, demonstrated that the states of the
ground-state manifold of this phase have a non-zero Hall conductivity, which is shown to be
precisely quantized and equal to the filling fraction, in units of $e^2/h$. The 
exact quantization of the Hall conductivity holds also when FCI states are
not ground states, and survives when the many-body Berry curvature is
not a particularly flat function. 

Deviating from the exact filling fraction $\nu=1/3$ by either removing electrons or increasing
the system size, we found indications of fractional quasihole
statistics in the eigenvalue spectra, by application of the
state-counting rule elaborated in Ref.~\onlinecite{Bernevig2012}. The eigenvalue
spectra contain features that remind one of both the hierarchy and
the composite fermion pictures of FQH states, therefore pushing the
correspondence between FCI and FQH states one step further. 


Since in the present context the magnetic
texture is generated by strongly correlated itinerant electrons,
imperfections in this texture are to be expected. We therefore investigate
the impact of a localized impurity that mixes states across momentum
sectors. Disorder causes splitting of the energy levels
corresponding to the degenerate FCI ground states, thus making these
states inequivalent in finite systems. However, the gap between these
states and the excited-state spectrum remains open and their average
Hall conductivity remains accurately fixed to the value of the filling
fraction, even though the Hall conductivities of the individual states
deviate from that value. This behavior illustrates the topological
protection of the FCI states. 

Finally, we address the robustness of FCI states with respect to band
flatness, interaction strength, filling and FS nesting. Very flat bands indeed turn out to
require the weakest interaction to produce an FCI state, corroborating 
the figure of merit $M$ [see Eq.~(\ref{eq:merit})] as a good
first step in assessing candidate systems. As soon as
one deviates from this limit, however, additional properties of the
system become crucial: in contrast to a perfectly flat LL, strong
particle-hole asymmetry can arise on a lattice, because some fillings
allow CDWs more easily than others. In addition, a CDW is favored by
FS nesting and this in turn reduces the stability range of the FCI
states. Nevertheless, it turns out that FCI states can be obtained
if either of the following conditions is met: (i) very flat bands, even near nesting and at a
filling favorable to a CDW, and (ii) even rather dispersive bands, if
the filling fraction or the absence of FS nesting are
unfavorable to a CDW. This flexibility makes FCI states appear more realistic.  

\begin{acknowledgments}
This work was supported by the Deutsche Forschungsgemeinschaft under
the Emmy-Noether program (S.K. and M.D.) and the Interphase Program of
the Dutch Science Foundation NWO/FOM (JV). We thank J. van den Brink for helpful discussions
and B.~A.~Bernevig for insightful comments.
\end{acknowledgments}

%


\begin{thebibliography}{58}%
\makeatletter
\providecommand \@ifxundefined [1]{%
 \@ifx{#1\undefined}
}%
\providecommand \@ifnum [1]{%
 \ifnum #1\expandafter \@firstoftwo
 \else \expandafter \@secondoftwo
 \fi
}%
\providecommand \@ifx [1]{%
 \ifx #1\expandafter \@firstoftwo
 \else \expandafter \@secondoftwo
 \fi
}%
\providecommand \natexlab [1]{#1}%
\providecommand \enquote  [1]{``#1''}%
\providecommand \bibnamefont  [1]{#1}%
\providecommand \bibfnamefont [1]{#1}%
\providecommand \citenamefont [1]{#1}%
\providecommand \href@noop [0]{\@secondoftwo}%
\providecommand \href [0]{\begingroup \@sanitize@url \@href}%
\providecommand \@href[1]{\@@startlink{#1}\@@href}%
\providecommand \@@href[1]{\endgroup#1\@@endlink}%
\providecommand \@sanitize@url [0]{\catcode `\\12\catcode `\$12\catcode
  `\&12\catcode `\#12\catcode `\^12\catcode `\_12\catcode `\%12\relax}%
\providecommand \@@startlink[1]{}%
\providecommand \@@endlink[0]{}%
\providecommand \url  [0]{\begingroup\@sanitize@url \@url }%
\providecommand \@url [1]{\endgroup\@href {#1}{\urlprefix }}%
\providecommand \urlprefix  [0]{URL }%
\providecommand \Eprint [0]{\href }%
\providecommand \doibase [0]{http://dx.doi.org/}%
\providecommand \selectlanguage [0]{\@gobble}%
\providecommand \bibinfo  [0]{\@secondoftwo}%
\providecommand \bibfield  [0]{\@secondoftwo}%
\providecommand \translation [1]{[#1]}%
\providecommand \BibitemOpen [0]{}%
\providecommand \bibitemStop [0]{}%
\providecommand \bibitemNoStop [0]{.\EOS\space}%
\providecommand \EOS [0]{\spacefactor3000\relax}%
\providecommand \BibitemShut  [1]{\csname bibitem#1\endcsname}%
\let\auto@bib@innerbib\@empty
\bibitem [{\citenamefont {Laughlin}(1983)}]{Laughlin1983}%
  \BibitemOpen
  \bibfield  {author} {\bibinfo {author} {\bibfnamefont {R.~B.}\ \bibnamefont
  {Laughlin}},\ }\href
  {\doibase 10.1103/PhysRevB.27.3383} {\bibfield  {journal} {\bibinfo  {journal}
  {Phys. Rev.~B}\ }\textbf {\bibinfo {volume} {27}},\ \bibinfo {pages}
  {3383} (\bibinfo {year} {1983})}\BibitemShut {NoStop}%
\bibitem [{\citenamefont {Haldane}(1983)}]{Haldane1983}%
  \BibitemOpen
  \bibfield  {author} {\bibinfo {author} {\bibfnamefont {F.~D.~M.}\
  \bibnamefont {Haldane}},\ }\href
  {\doibase 10.1103/PhysRevLett.51.605} {\bibfield  {journal} {\bibinfo
  {journal} {Phys. Rev. Lett.}\ }\textbf {\bibinfo {volume} {51}},\
  \bibinfo {pages} {605} (\bibinfo {year} {1983})}\BibitemShut {NoStop}%
\bibitem [{\citenamefont {Halperin}(1984)}]{Halperin1984}%
  \BibitemOpen
  \bibfield  {author} {\bibinfo {author} {\bibfnamefont {B.~I.}\ \bibnamefont
  {Halperin}},\ }\href
  {\doibase 10.1103/PhysRevLett.52.1583} {\bibfield  {journal} {\bibinfo  {journal}
  {Phys. Rev. Lett.}\ }\textbf {\bibinfo {volume} {52}},\ \bibinfo
  {pages} {1583} (\bibinfo {year} {1984})}\BibitemShut {NoStop}%
\bibitem [{\citenamefont {Tsui}\ \emph {et~al.}(1982)\citenamefont {Tsui},
  \citenamefont {Stormer},\ and\ \citenamefont {Gossard}}]{Tsui1982}%
  \BibitemOpen
  \bibfield  {author} {\bibinfo {author} {\bibfnamefont {D.~C.}\ \bibnamefont
  {Tsui}}, \bibinfo {author} {\bibfnamefont {H.~L.}\ \bibnamefont {Stormer}}, \
  and\ \bibinfo {author} {\bibfnamefont {A.~C.}\ \bibnamefont {Gossard}},\
  }\href
  {\doibase 10.1103/PhysRevLett.48.1559} {\bibfield  {journal} {\bibinfo  {journal} {Phys. Rev.  Lett.}\ }\textbf {\bibinfo {volume} {48}},\ \bibinfo {pages} {1559}
  (\bibinfo {year} {1982})}\BibitemShut {NoStop}%
\bibitem [{\citenamefont {Wilczek}(1982)}]{Wilczek1982}%
  \BibitemOpen
  \bibfield  {author} {\bibinfo {author} {\bibfnamefont {F.}~\bibnamefont
  {Wilczek}},\ }\href
  {\doibase 10.1103/PhysRevLett.49.957} {\bibfield  {journal} {\bibinfo  {journal}
  {Phys. Rev. Lett.}\ }\textbf {\bibinfo {volume} {49}},\ \bibinfo {pages} {957}
  (\bibinfo {year} {1982})}\BibitemShut {NoStop}%
\bibitem [{\citenamefont {Haldane}(1991)}]{Haldane1991}%
  \BibitemOpen
  \bibfield  {author} {\bibinfo {author} {\bibfnamefont {F.~D.~M.}\
  \bibnamefont {Haldane}},\ }\href
  {\doibase 10.1103/PhysRevLett.67.937} {\bibfield  {journal} {\bibinfo
  {journal} {Phys. Rev. Lett.}\ }\textbf {\bibinfo {volume} {67}},\
  \bibinfo {pages} {937} (\bibinfo {year} {1991})}\BibitemShut {NoStop}%
\bibitem [{\citenamefont {Nayak}\ \emph {et~al.}(2008)\citenamefont {Nayak},
  \citenamefont {Stern}, \citenamefont {Freedman},\ and\ \citenamefont {{Das
  Sarma}}}]{Nayak2008}%
  \BibitemOpen
  \bibfield  {author} {\bibinfo {author} {\bibfnamefont {C.}~\bibnamefont
  {Nayak}}, \bibinfo {author} {\bibfnamefont {A.}~\bibnamefont {Stern}},
  \bibinfo {author} {\bibfnamefont {M.}~\bibnamefont {Freedman}}, \ and\
  \bibinfo {author} {\bibfnamefont {S.}~\bibnamefont {{Das Sarma}}},\ }\href
  {\doibase 10.1103/RevModPhys.80.1083} {\bibfield  {journal} {\bibinfo
  {journal} {Rev. Mod. Phys.}\ }\textbf {\bibinfo {volume} {80}},\
  \bibinfo {pages} {1083} (\bibinfo {year} {2008})}\BibitemShut {NoStop}%
\bibitem [{\citenamefont {Hofstadter}(1976)}]{Hofstadter1976}%
  \BibitemOpen
  \bibfield  {author} {\bibinfo {author} {\bibfnamefont {D.~R.}\ \bibnamefont
  {Hofstadter}},\ }\href
  {\doibase 10.1103/PhysRevB.14.2239} {\bibfield  {journal} {\bibinfo  {journal}
  {Phys. Rev.~B}\ }\textbf {\bibinfo {volume} {14}},\ \bibinfo {pages}
  {2239} (\bibinfo {year} {1976})}\BibitemShut {NoStop}%
\bibitem [{\citenamefont {Kliros}\ and\ \citenamefont
  {D'Ambrumenil}(1991)}]{Kliros1991}%
  \BibitemOpen
  \bibfield  {author} {\bibinfo {author} {\bibfnamefont {G.~S.}\ \bibnamefont
  {Kliros}}\ and\ \bibinfo {author} {\bibfnamefont {N.}~\bibnamefont
  {D'Ambrumenil}},\ }\href
  {\doibase 10.1088/0953-8984/3/23/012} {\bibfield  {journal} {\bibinfo  {journal}
  {J. Phys.~C: Solid State Phys.}\ }\textbf {\bibinfo {volume}
  {3}},\ \bibinfo {pages} {4241} (\bibinfo {year} {1991})}\BibitemShut
  {NoStop}%
\bibitem [{\citenamefont {Haldane}(1988)}]{haldane1988}%
  \BibitemOpen
  \bibfield  {author} {\bibinfo {author} {\bibfnamefont {F.~D.~M.}\
  \bibnamefont {Haldane}},\ }\href {\doibase 10.1103/PhysRevLett.61.2015}
  {\bibfield  {journal} {\bibinfo  {journal} {Phys. Rev. Lett.}\
  }\textbf {\bibinfo {volume} {61}},\ \bibinfo {pages} {2015} (\bibinfo {year}
  {1988})}\BibitemShut {NoStop}%
\bibitem [{\citenamefont {Qi}\ \emph {et~al.}(2006)\citenamefont {Qi},
  \citenamefont {Wu},\ and\ \citenamefont {Zhang}}]{Qi:2006jm}%
  \BibitemOpen
  \bibfield  {author} {\bibinfo {author} {\bibfnamefont {X.-L.}\ \bibnamefont
  {Qi}}, \bibinfo {author} {\bibfnamefont {Y.-S.}\ \bibnamefont {Wu}}, \ and\
  \bibinfo {author} {\bibfnamefont {S.-C.}\ \bibnamefont {Zhang}},\ }\href
  {\doibase 10.1103/PhysRevB.74.085308} {\bibfield  {journal} {\bibinfo  {journal} {Phys. Rev.~B}\ }\textbf
  {\bibinfo {volume} {74}},\ \bibinfo {pages} {085308} (\bibinfo {year}
  {2006})}\BibitemShut {NoStop}%
\bibitem [{\citenamefont {Liu}\ \emph {et~al.}(2008)\citenamefont {Liu},
  \citenamefont {Qi}, \citenamefont {Dai}, \citenamefont {Fang},\ and\
  \citenamefont {Zhang}}]{Liu:2008ev}%
  \BibitemOpen
  \bibfield  {author} {\bibinfo {author} {\bibfnamefont {C.-X.}\ \bibnamefont
  {Liu}}, \bibinfo {author} {\bibfnamefont {X.-L.}\ \bibnamefont {Qi}},
  \bibinfo {author} {\bibfnamefont {X.}~\bibnamefont {Dai}}, \bibinfo {author}
  {\bibfnamefont {Z.}~\bibnamefont {Fang}}, \ and\ \bibinfo {author}
  {\bibfnamefont {S.-C.}\ \bibnamefont {Zhang}},\ }\href
  {\doibase 10.1103/PhysRevLett.101.146802} {\bibfield
  {journal} {\bibinfo  {journal} {Phys. Rev. Lett.}\ }\textbf {\bibinfo
  {volume} {101}},\ \bibinfo {pages} {146802} (\bibinfo {year}
  {2008})}\BibitemShut {NoStop}%
\bibitem [{\citenamefont {Ohgushi}\ \emph {et~al.}(2000)\citenamefont
  {Ohgushi}, \citenamefont {Murakami},\ and\ \citenamefont
  {Nagaosa}}]{Ohgushi2000}%
  \BibitemOpen
  \bibfield  {author} {\bibinfo {author} {\bibfnamefont {K.}~\bibnamefont
  {Ohgushi}}, \bibinfo {author} {\bibfnamefont {S.}~\bibnamefont {Murakami}}, \
  and\ \bibinfo {author} {\bibfnamefont {N.}~\bibnamefont {Nagaosa}},\
  }\href
  {\doibase 10.1103/PhysRevB.62.6065} {\bibfield  {journal} {\bibinfo  {journal} {Phys. Rev. 
  B}\ }\textbf {\bibinfo {volume} {62}},\ \bibinfo {pages} {6065} (\bibinfo
  {year} {2000})}\BibitemShut {NoStop}%
\bibitem [{\citenamefont {Martin}\ and\ \citenamefont
  {Batista}(2008)}]{Martin2008}%
  \BibitemOpen
  \bibfield  {author} {\bibinfo {author} {\bibfnamefont {I.}~\bibnamefont
  {Martin}}\ and\ \bibinfo {author} {\bibfnamefont {C.~D.}\ \bibnamefont
  {Batista}},\ }\href {\doibase 10.1103/PhysRevLett.101.156402} {\bibfield
  {journal} {\bibinfo  {journal} {Phys. Rev. Lett.}\ }\textbf {\bibinfo
  {volume} {101}},\ \bibinfo {pages} {156402} (\bibinfo {year}
  {2008})}\BibitemShut {NoStop}%
\bibitem [{\citenamefont {Tang}\ \emph {et~al.}(2011)\citenamefont {Tang},
  \citenamefont {Mei},\ and\ \citenamefont {Wen}}]{Tang2011}%
  \BibitemOpen
  \bibfield  {author} {\bibinfo {author} {\bibfnamefont {E.}~\bibnamefont
  {Tang}}, \bibinfo {author} {\bibfnamefont {J.-W.}\ \bibnamefont {Mei}}, \
  and\ \bibinfo {author} {\bibfnamefont {X.-G.}\ \bibnamefont {Wen}},\ }\href
  {\doibase 10.1103/PhysRevLett.106.236802} {\bibfield  {journal} {\bibinfo
  {journal} {Phys. Rev. Lett.}\ }\textbf {\bibinfo {volume} {106}},\
  \bibinfo {pages} {236802} (\bibinfo {year} {2011})}\BibitemShut {NoStop}%
\bibitem [{\citenamefont {Sun}\ \emph {et~al.}(2011)\citenamefont {Sun},
  \citenamefont {Gu}, \citenamefont {Katsura},\ and\ \citenamefont {{Das
  Sarma}}}]{Sun2011}%
  \BibitemOpen
  \bibfield  {author} {\bibinfo {author} {\bibfnamefont {K.}~\bibnamefont
  {Sun}}, \bibinfo {author} {\bibfnamefont {Z.}~\bibnamefont {Gu}}, \bibinfo
  {author} {\bibfnamefont {H.}~\bibnamefont {Katsura}}, \ and\ \bibinfo
  {author} {\bibfnamefont {S.}~\bibnamefont {{Das Sarma}}},\ }\href {\doibase
  10.1103/PhysRevLett.106.236803} {\bibfield  {journal} {\bibinfo  {journal}
  {Phys. Rev. Lett.}\ }\textbf {\bibinfo {volume} {106}},\ \bibinfo
  {pages} {236803} (\bibinfo {year} {2011})} \BibitemShut {NoStop}%
\bibitem [{\citenamefont {Neupert}\ \emph {et~al.}(2011)\citenamefont
  {Neupert}, \citenamefont {Santos}, \citenamefont {Chamon},\ and\
  \citenamefont {Mudry}}]{Neupert2011}%
  \BibitemOpen
  \bibfield  {author} {\bibinfo {author} {\bibfnamefont {T.}~\bibnamefont
  {Neupert}}, \bibinfo {author} {\bibfnamefont {L.}~\bibnamefont {Santos}},
  \bibinfo {author} {\bibfnamefont {C.}~\bibnamefont {Chamon}}, \ and\ \bibinfo
  {author} {\bibfnamefont {C.}~\bibnamefont {Mudry}},\ }\href {\doibase
  10.1103/PhysRevLett.106.236804} {\bibfield  {journal} {\bibinfo  {journal}
  {Phys. Rev. Lett.}\ }\textbf {\bibinfo {volume} {106}},\ \bibinfo
  {pages} {236804} (\bibinfo {year} {2011})}\BibitemShut {NoStop}%
\bibitem [{\citenamefont {Sheng}\ \emph {et~al.}(2011)\citenamefont {Sheng},
  \citenamefont {Gu}, \citenamefont {Sun},\ and\ \citenamefont
  {Sheng}}]{Sheng2011}%
  \BibitemOpen
  \bibfield  {author} {\bibinfo {author} {\bibfnamefont {D.~N.}\ \bibnamefont
  {Sheng}}, \bibinfo {author} {\bibfnamefont {Z.-C.}\ \bibnamefont {Gu}},
  \bibinfo {author} {\bibfnamefont {K.}~\bibnamefont {Sun}}, \ and\ \bibinfo
  {author} {\bibfnamefont {L.}~\bibnamefont {Sheng}},\ }\href {\doibase
  10.1038/ncomms1380} {\bibfield  {journal} {\bibinfo  {journal} {Nat. 
  Comm.}\ }\textbf {\bibinfo {volume} {2}},\ \bibinfo {pages} {389}
  (\bibinfo {year} {2011})}\BibitemShut {NoStop}%
\bibitem [{\citenamefont {Wang}\ \emph {et~al.}(2011)\citenamefont {Wang},
  \citenamefont {Gu}, \citenamefont {Gong},\ and\ \citenamefont
  {Sheng}}]{Wang2011}%
  \BibitemOpen
  \bibfield  {author} {\bibinfo {author} {\bibfnamefont {Y.-F.}\ \bibnamefont
  {Wang}}, \bibinfo {author} {\bibfnamefont {Z.-C.}\ \bibnamefont {Gu}},
  \bibinfo {author} {\bibfnamefont {C.-D.}\ \bibnamefont {Gong}}, \ and\
  \bibinfo {author} {\bibfnamefont {D. N.}~\bibnamefont {Sheng}},\ }\href
  {\doibase 10.1103/PhysRevLett.107.146803} {\bibfield  {journal} {\bibinfo
  {journal} {Phys. Rev. Lett.}\ }\textbf {\bibinfo {volume} {107}},\
  \bibinfo {pages} {146803} (\bibinfo {year} {2011})}\BibitemShut {NoStop}%
\bibitem [{\citenamefont {Regnault}\ and\ \citenamefont
  {Bernevig}(2011)}]{Regnault2011}%
  \BibitemOpen
  \bibfield  {author} {\bibinfo {author} {\bibfnamefont {N.}~\bibnamefont
  {Regnault}}\ and\ \bibinfo {author} {\bibfnamefont {B.~A.}\ \bibnamefont
  {Bernevig}},\ }\href
  {\doibase 10.1103/PhysRevX.1.021014}{\bibfield  {journal} {\bibinfo  {journal}
  {Phys. Rev.  X}\ }\textbf {\bibinfo {volume} {1}},\ \bibinfo {pages}
  {021014} (\bibinfo {year} {2011})} \BibitemShut {NoStop}%
\bibitem [{\citenamefont {Wu}\ \emph {et~al.}(2012{\natexlab{a}})\citenamefont
  {Wu}, \citenamefont {Bernevig},\ and\ \citenamefont {Regnault}}]{Wu2012}%
  \BibitemOpen
  \bibfield  {author} {\bibinfo {author} {\bibfnamefont {Y.~L.}\ \bibnamefont
  {Wu}}, \bibinfo {author} {\bibfnamefont {B.~A.}\ \bibnamefont {Bernevig}}, \
  and\ \bibinfo {author} {\bibfnamefont {N.}~\bibnamefont {Regnault}},\
  }\href
  {\doibase 10.1103/PhysRevB.85.075116}{\bibfield  {journal} {\bibinfo  {journal} {Phys. Rev. 
  B}\ }\textbf {\bibinfo {volume} {85}},\ \bibinfo {pages} {075116} (\bibinfo
  {year} {2012}{\natexlab{a}})} \BibitemShut {NoStop}%
\bibitem [{\citenamefont {Venderbos}\ \emph {et~al.}(2012)\citenamefont
  {Venderbos}, \citenamefont {Kourtis}, \citenamefont {van~den Brink},\ and\
  \citenamefont {Daghofer}}]{Venderbos2012}%
  \BibitemOpen
  \bibfield  {author} {\bibinfo {author} {\bibfnamefont {J.~W.~F.}\
  \bibnamefont {Venderbos}}, \bibinfo {author} {\bibfnamefont {S.}~\bibnamefont
  {Kourtis}}, \bibinfo {author} {\bibfnamefont {J.}~\bibnamefont {van~den
  Brink}}, \ and\ \bibinfo {author} {\bibfnamefont {M.}~\bibnamefont
  {Daghofer}},\ }\href{http://link.aps.org/doi/10.1103/PhysRevLett.108.126405} {\bibfield  {journal} {\bibinfo  {journal}
  {Phys. Rev. Lett.}\ }\textbf {\bibinfo {volume} {108}},\ \bibinfo {pages}
  {126405} (\bibinfo {year} {2012})} \BibitemShut
  {NoStop}%
\bibitem [{\citenamefont {Liu}\ \emph {et~al.}(2012{\natexlab{a}})\citenamefont
  {Liu}, \citenamefont {Repellin}, \citenamefont {Bernevig},\ and\
  \citenamefont {Regnault}}]{Liu2012a}%
  \BibitemOpen
  \bibfield  {author} {\bibinfo {author} {\bibfnamefont {T.}~\bibnamefont
  {Liu}}, \bibinfo {author} {\bibfnamefont {C.}~\bibnamefont {Repellin}},
  \bibinfo {author} {\bibfnamefont {B.~A.}\ \bibnamefont {Bernevig}}, \ and\
  \bibinfo {author} {\bibfnamefont {N.}~\bibnamefont {Regnault}},\ }\href@noop
  {} \Eprint {http://arxiv.org/abs/1206.2626v1} {arXiv:1206.2626v1 (2012)} \BibitemShut {NoStop}%
\bibitem [{\citenamefont {{L{\"a}uchli}}\ \emph {et~al.}(2012)\citenamefont
  {{L{\"a}uchli}}, \citenamefont {{Liu}}, \citenamefont {{Bergholtz}},\ and\
  \citenamefont {{Moessner}}}]{2012arXiv1207.6094L}%
  \BibitemOpen
  \bibfield  {author} {\bibinfo {author} {\bibfnamefont {A.~M.}\ \bibnamefont
  {{L{\"a}uchli}}}, \bibinfo {author} {\bibfnamefont {Z.}~\bibnamefont
  {{Liu}}}, \bibinfo {author} {\bibfnamefont {E.~J.}\ \bibnamefont
  {{Bergholtz}}}, \ and\ \bibinfo {author} {\bibfnamefont {R.}~\bibnamefont
  {{Moessner}}},\ }\href@noop {} \Eprint
  {http://arxiv.org/abs/1207.6094v1} {arXiv:1207.6094v1 (2012)}
  \BibitemShut {NoStop}%
\bibitem [{\citenamefont {Trescher}\ and\ \citenamefont
  {Bergholtz}(2012)}]{Trescher2012}%
  \BibitemOpen
  \bibfield  {author} {\bibinfo {author} {\bibfnamefont {M.}~\bibnamefont
  {Trescher}}\ and\ \bibinfo {author} {\bibfnamefont {E.~J.}~\bibnamefont
  {Bergholtz}},\ }\href@noop {} \Eprint
  {http://arxiv.org/abs/arXiv:1205.2245v3} {arXiv:1205.2245v3 (2012)}
  \BibitemShut {NoStop}%
\bibitem [{\citenamefont {Yang}\ \emph
  {et~al.}(2012{\natexlab{a}})\citenamefont {Yang}, \citenamefont {Gu},
  \citenamefont {Sun},\ and\ \citenamefont {Sarma}}]{Yang2012a}%
  \BibitemOpen
  \bibfield  {author} {\bibinfo {author} {\bibfnamefont {S.}~\bibnamefont
  {Yang}}, \bibinfo {author} {\bibfnamefont {Z.-C.}\ \bibnamefont {Gu}},
  \bibinfo {author} {\bibfnamefont {K.}~\bibnamefont {Sun}}, \ and\ \bibinfo
  {author} {\bibfnamefont {S.~D.}\ \bibnamefont {Sarma}},\ }\href@noop{} \Eprint
  {http://arxiv.org/abs/arXiv:1205.5792v2} {arXiv:1205.5792v2 (2012)}
  \BibitemShut {NoStop}%
\bibitem [{\citenamefont {Wang}\ \emph {et~al.}(2012)\citenamefont {Wang},
  \citenamefont {Yao}, \citenamefont {Gong},\ and\ \citenamefont
  {Sheng}}]{Wang2012}%
  \BibitemOpen
  \bibfield  {author} {\bibinfo {author} {\bibfnamefont {Y.-F.}\ \bibnamefont
  {Wang}}, \bibinfo {author} {\bibfnamefont {H.}~\bibnamefont {Yao}}, \bibinfo
  {author} {\bibfnamefont {C.-D.}\ \bibnamefont {Gong}}, \ and\ \bibinfo
  {author} {\bibfnamefont {D.~N.}\ \bibnamefont {Sheng}},\ }\href{http://link.aps.org/doi/10.1103/PhysRevB.86.201101} {\bibfield  {journal} {\bibinfo  {journal}
  {Phys. Rev.~B}\ }\textbf {\bibinfo {volume} {86}},\ \bibinfo {pages}
  {201101(R)} (\bibinfo {year} {2012})} \BibitemShut
  {NoStop}%
  {arXiv:1204.1697v1 (2012)} \BibitemShut {NoStop}%
\bibitem [{\citenamefont {Liu}\ \emph {et~al.}(2012{\natexlab{b}})\citenamefont
  {Liu}, \citenamefont {Bergholtz}, \citenamefont {Fan},\ and\ \citenamefont
  {L\"{a}uchli}}]{Liu2012b}%
  \BibitemOpen
  \bibfield  {author} {\bibinfo {author} {\bibfnamefont {Z.}~\bibnamefont
  {Liu}}, \bibinfo {author} {\bibfnamefont {E.~J.}\ \bibnamefont {Bergholtz}},
  \bibinfo {author} {\bibfnamefont {H.}~\bibnamefont {Fan}}, \ and\ \bibinfo
  {author} {\bibfnamefont {A.}~\bibnamefont {L\"{a}uchli}},\ }\href{http://link.aps.org/doi/10.1103/PhysRevLett.109.186805} {\bibfield  {journal} {\bibinfo  {journal}
  {Phys. Rev. Lett.}\ }\textbf {\bibinfo {volume} {109}},\ \bibinfo {pages}
  {186805} (\bibinfo {year} {2012})} \BibitemShut
  {NoStop}%
\bibitem [{\citenamefont {Sterdyniak}\ \emph {et~al.}(2012)\citenamefont
  {Sterdyniak}, \citenamefont {Repellin}, \citenamefont {Bernevig},\ and\
  \citenamefont {Regnault}}]{Sterdyniak2012}%
  \BibitemOpen
  \bibfield  {author} {\bibinfo {author} {\bibfnamefont {A.}~\bibnamefont
  {Sterdyniak}}, \bibinfo {author} {\bibfnamefont {C.}~\bibnamefont
  {Repellin}}, \bibinfo {author} {\bibfnamefont {B.~A.}\ \bibnamefont
  {Bernevig}}, \ and\ \bibinfo {author} {\bibfnamefont {N.}~\bibnamefont
  {Regnault}},\ }\href@noop {} \Eprint
  {http://arxiv.org/abs/arXiv:1207.6385v1} {arXiv:1207.6385v1 (2012)}
  \BibitemShut {NoStop}%
\bibitem [{\citenamefont {{Grushin}}\ \emph {et~al.}(2012)\citenamefont
  {{Grushin}}, \citenamefont {{Neupert}}, \citenamefont {{Chamon}},\ and\
  \citenamefont {{Mudry}}}]{2012arXiv1207.4097G}%
  \BibitemOpen
  \bibfield  {author} {\bibinfo {author} {\bibfnamefont {A.~G.}\ \bibnamefont
  {{Grushin}}}, \bibinfo {author} {\bibfnamefont {T.}~\bibnamefont
  {{Neupert}}}, \bibinfo {author} {\bibfnamefont {C.}~\bibnamefont {{Chamon}}},
  \ and\ \bibinfo {author} {\bibfnamefont {C.}~\bibnamefont {{Mudry}}},\ }\href{http://link.aps.org/doi/10.1103/PhysRevB.86.205125} {\bibfield  {journal} {\bibinfo  {journal}
  {Phys. Rev.~B}\ }\textbf {\bibinfo {volume} {86}},\ \bibinfo {pages}
  {205125} (\bibinfo {year} {2012})} \BibitemShut
  {NoStop}%
\bibitem [{\citenamefont {Bernevig}\ and\ \citenamefont
  {Regnault}(2012)}]{Bernevig2012}%
  \BibitemOpen
  \bibfield  {author} {\bibinfo {author} {\bibfnamefont {B.~A.}\ \bibnamefont
  {Bernevig}}\ and\ \bibinfo {author} {\bibfnamefont {N.}~\bibnamefont
  {Regnault}},\ }\href
  {\doibase 10.1103/PhysRevB.85.075128}{\bibfield  {journal} {\bibinfo  {journal}
  {Phys. Rev.~B}\ }\textbf {\bibinfo {volume} {85}},\ \bibinfo {pages}
  {075128} (\bibinfo {year} {2012})} \BibitemShut {NoStop}%
\bibitem [{\citenamefont {Qi}(2011)}]{Qi2011}%
  \BibitemOpen
  \bibfield  {author} {\bibinfo {author} {\bibfnamefont {X.-L.}\ \bibnamefont
  {Qi}},\ }\href
  {\doibase 10.1103/PhysRevLett.107.126803}{\bibfield  {journal}
  {\bibinfo  {journal} {Phys. Rev. Lett.}\ }\textbf {\bibinfo {volume}
  {107}},\ \bibinfo {pages} {126803} (\bibinfo {year} {2011})} \BibitemShut {NoStop}%
\bibitem [{\citenamefont {Wu}\ \emph {et~al.}(2012{\natexlab{b}})\citenamefont
  {Wu}, \citenamefont {Regnault},\ and\ \citenamefont {Bernevig}}]{Wu2012b}%
  \BibitemOpen
  \bibfield  {author} {\bibinfo {author} {\bibfnamefont {Y.-L.}\ \bibnamefont
  {Wu}}, \bibinfo {author} {\bibfnamefont {N.}~\bibnamefont {Regnault}}, \ and\
  \bibinfo {author} {\bibfnamefont {B.~A.}\ \bibnamefont {Bernevig}},\
  }\href{http://link.aps.org/doi/10.1103/PhysRevB.86.085129} {\bibfield  {journal} {\bibinfo  {journal}
  {Phys. Rev.~B}\ }\textbf {\bibinfo {volume} {86}},\ \bibinfo {pages}
  {085129} (\bibinfo {year} {2012})} \BibitemShut
  {NoStop}%
\bibitem [{\citenamefont {Lee}\ \emph {et~al.}(2012)\citenamefont {Lee},
  \citenamefont {Thomale},\ and\ \citenamefont {Qi}}]{Lee2012}%
  \BibitemOpen
  \bibfield  {author} {\bibinfo {author} {\bibfnamefont {C.~H.}\ \bibnamefont
  {Lee}}, \bibinfo {author} {\bibfnamefont {R.}~\bibnamefont {Thomale}}, \ and\
  \bibinfo {author} {\bibfnamefont {X.-L.}\ \bibnamefont {Qi}},\ }\href@noop {}
  \Eprint {http://arxiv.org/abs/arXiv:1207.5587v2}
  {arXiv:1207.5587v2 (2012)} \BibitemShut {NoStop}%
\bibitem [{\citenamefont {Murthy}\ and\ \citenamefont
  {Shankar}(2012)}]{Murthy2012}%
  \BibitemOpen
  \bibfield  {author} {\bibinfo {author} {\bibfnamefont {G.}~\bibnamefont
  {Murthy}}\ and\ \bibinfo {author} {\bibfnamefont {R.}~\bibnamefont
  {Shankar}},\ }\href@noop {} \Eprint
  {http://arxiv.org/abs/arXiv:1207.2133v2} {arXiv:1207.2133v2 (2012)}
  \BibitemShut {NoStop}%
\bibitem [{\citenamefont {Murthy}\ and\ \citenamefont
  {Shankar}(2011)}]{Murthy2011}%
  \BibitemOpen
  \bibfield  {author} {\bibinfo {author} {\bibfnamefont {G.}~\bibnamefont
  {Murthy}}\ and\ \bibinfo {author} {\bibfnamefont {R.}~\bibnamefont
  {Shankar}},\ }\href@noop {} \Eprint
  {http://arxiv.org/abs/arXiv:1108.5501v2} {arXiv:1108.5501v2 (2012)}
  \BibitemShut {NoStop}%
\bibitem [{\citenamefont {Parameswaran}\ \emph {et~al.}(2012)\citenamefont
  {Parameswaran}, \citenamefont {Roy},\ and\ \citenamefont
  {Sondhi}}]{Parameswaran2012}%
  \BibitemOpen
  \bibfield  {author} {\bibinfo {author} {\bibfnamefont {S.~A.}\ \bibnamefont
  {Parameswaran}}, \bibinfo {author} {\bibfnamefont {R.}~\bibnamefont {Roy}}, \
  and\ \bibinfo {author} {\bibfnamefont {S.~L.}\ \bibnamefont {Sondhi}},\
  }\href {\doibase 10.1103/PhysRevB.85.241308} {\bibfield  {journal} {\bibinfo
  {journal} {Phys. Rev.~B}\ }\textbf {\bibinfo {volume} {85}},\ \bibinfo
  {pages} {241308} (\bibinfo {year} {2012})}\BibitemShut {NoStop}%
\bibitem [{\citenamefont {Goerbig}(2012)}]{Goerbig2012}%
  \BibitemOpen
  \bibfield  {author} {\bibinfo {author} {\bibfnamefont {M.~O.}\ \bibnamefont
  {Goerbig}},\ }\href {\doibase 10.1140/epjb/e2011-20857-6} {\bibfield
  {journal} {\bibinfo  {journal} {Eur. Phys. J.~B}\ }\textbf
  {\bibinfo {volume} {85}},\ \bibinfo {pages} {15} (\bibinfo {year}
  {2012})}\BibitemShut {NoStop}%
\bibitem [{\citenamefont {Xiao}\ \emph {et~al.}(2011)\citenamefont {Xiao},
  \citenamefont {Zhu}, \citenamefont {Ran}, \citenamefont {Nagaosa},\ and\
  \citenamefont {Okamoto}}]{Xiao:2011}%
  \BibitemOpen
  \bibfield  {author} {\bibinfo {author} {\bibfnamefont {D.}~\bibnamefont
  {Xiao}}, \bibinfo {author} {\bibfnamefont {W.}~\bibnamefont {Zhu}}, \bibinfo
  {author} {\bibfnamefont {Y.}~\bibnamefont {Ran}}, \bibinfo {author}
  {\bibfnamefont {N.}~\bibnamefont {Nagaosa}}, \ and\ \bibinfo {author}
  {\bibfnamefont {S.}~\bibnamefont {Okamoto}},\ }\href {\doibase
  10.1038/ncomms1602} {\bibfield  {journal} {\bibinfo  {journal} {Nat. 
  Comm.}\ }\textbf {\bibinfo {volume} {2}},\ \bibinfo {pages} {596}
  (\bibinfo {year} {2011})}\BibitemShut {NoStop}%
\bibitem [{\citenamefont {Ghaemi}\ \emph {et~al.}(2012)\citenamefont {Ghaemi},
  \citenamefont {Cayssol}, \citenamefont {Sheng},\ and\ \citenamefont
  {Vishwanath}}]{PhysRevLett.108.266801}%
  \BibitemOpen
  \bibfield  {author} {\bibinfo {author} {\bibfnamefont {P.}~\bibnamefont
  {Ghaemi}}, \bibinfo {author} {\bibfnamefont {J.}~\bibnamefont {Cayssol}},
  \bibinfo {author} {\bibfnamefont {D.~N.}\ \bibnamefont {Sheng}}, \ and\
  \bibinfo {author} {\bibfnamefont {A.}~\bibnamefont {Vishwanath}},\ }\href
  {\doibase 10.1103/PhysRevLett.108.266801} {\bibfield  {journal} {\bibinfo
  {journal} {Phys. Rev. Lett.}\ }\textbf {\bibinfo {volume} {108}},\ \bibinfo
  {pages} {266801} (\bibinfo {year} {2012})}\BibitemShut {NoStop}%
\bibitem [{\citenamefont {Venderbos}\ \emph {et~al.}(2011)\citenamefont
  {Venderbos}, \citenamefont {Daghofer},\ and\ \citenamefont {van~den
  Brink}}]{Venderbos2011}%
  \BibitemOpen
  \bibfield  {author} {\bibinfo {author} {\bibfnamefont {J.~W.~F.}\
  \bibnamefont {Venderbos}}, \bibinfo {author} {\bibfnamefont {M.}~\bibnamefont
  {Daghofer}}, \ and\ \bibinfo {author} {\bibfnamefont {J.}~\bibnamefont
  {van~den Brink}},\ }\href {http://arxiv.org/abs/1106.4439} {\bibfield
  {journal} {\bibinfo  {journal} {Phys. Rev. Lett.}\ }\textbf {\bibinfo
  {volume} {107}},\ \bibinfo {pages} {116401} (\bibinfo {year}
{2011})} \BibitemShut
  {NoStop}%
\bibitem [{\citenamefont {Akagi}\ and\ \citenamefont
  {Motome}(2010)}]{Akagi:2010p083711}%
  \BibitemOpen
  \bibfield  {author} {\bibinfo {author} {\bibfnamefont {Y.}~\bibnamefont
  {Akagi}}\ and\ \bibinfo {author} {\bibfnamefont {Y.}~\bibnamefont {Motome}},\
  }\href {\doibase info:doi/10.1143/JPSJ.79.083711} {\bibfield  {journal}
  {\bibinfo  {journal} {J. Phys. Soc. Jpn.}\ }\textbf {\bibinfo {volume}
  {79}},\ \bibinfo {pages} {083711} (\bibinfo {year} {2010})}\BibitemShut
  {NoStop}%
\bibitem [{\citenamefont {Kumar}\ and\ \citenamefont {{van den
  Brink}}(2010)}]{Kumar:2010p216405}%
  \BibitemOpen
  \bibfield  {author} {\bibinfo {author} {\bibfnamefont {S.}~\bibnamefont
  {Kumar}}\ and\ \bibinfo {author} {\bibfnamefont {J.}~\bibnamefont {{van den
  Brink}}},\ }\href {\doibase 10.1103/PhysRevLett.105.216405} {\bibfield
  {journal} {\bibinfo  {journal} {Phys.~Rev.~Lett.}\ }\textbf {\bibinfo
  {volume} {105}},\ \bibinfo {pages} {216405} (\bibinfo {year}
  {2010})}\BibitemShut {NoStop}%
\bibitem [{\citenamefont {Kato}\ \emph {et~al.}(2010)\citenamefont {Kato},
  \citenamefont {Martin},\ and\ \citenamefont {Batista}}]{Kato_FKLM_tri_2010}%
  \BibitemOpen
  \bibfield  {author} {\bibinfo {author} {\bibfnamefont {Y.}~\bibnamefont
  {Kato}}, \bibinfo {author} {\bibfnamefont {I.}~\bibnamefont {Martin}}, \ and\
  \bibinfo {author} {\bibfnamefont {C.~D.}\ \bibnamefont {Batista}},\ }\href
  {\doibase 10.1103/PhysRevLett.105.266405} {\bibfield  {journal} {\bibinfo
  {journal} {Phys. Rev. Lett.}\ }\textbf {\bibinfo {volume} {105}},\ \bibinfo
  {pages} {266405} (\bibinfo {year} {2010})}\BibitemShut {NoStop}%
\bibitem [{\citenamefont {Pen}\ \emph {et~al.}(1997)\citenamefont {Pen},
  \citenamefont {van~den Brink}, \citenamefont {Khomskii},\ and\ \citenamefont
  {Sawatzky}}]{Pen97}%
  \BibitemOpen
  \bibfield  {author} {\bibinfo {author} {\bibfnamefont {H.~F.}\ \bibnamefont
  {Pen}}, \bibinfo {author} {\bibfnamefont {J.}~\bibnamefont {van~den Brink}},
  \bibinfo {author} {\bibfnamefont {D.~I.}\ \bibnamefont {Khomskii}}, \ and\
  \bibinfo {author} {\bibfnamefont {G.~A.}\ \bibnamefont {Sawatzky}},\ }\href
  {\doibase 10.1103/PhysRevLett.78.1323} {\bibfield  {journal} {\bibinfo
  {journal} {Phys. Rev. Lett.}\ }\textbf {\bibinfo {volume} {78}},\ \bibinfo
  {pages} {1323} (\bibinfo {year} {1997})}\BibitemShut {NoStop}%
\bibitem [{\citenamefont {Koshibae}\ and\ \citenamefont
  {Maekawa}(2003)}]{Koshibae03}%
  \BibitemOpen
  \bibfield  {author} {\bibinfo {author} {\bibfnamefont {W.}~\bibnamefont
  {Koshibae}}\ and\ \bibinfo {author} {\bibfnamefont {S.}~\bibnamefont
  {Maekawa}},\ }\href {\doibase 10.1103/PhysRevLett.91.257003} {\bibfield
  {journal} {\bibinfo  {journal} {Phys. Rev. Lett.}\ }\textbf {\bibinfo
  {volume} {91}},\ \bibinfo {pages} {257003} (\bibinfo {year}
  {2003})}\BibitemShut {NoStop}%
  \bibitem [{\citenamefont {Harrison}(1980)}]{har80}%
  \BibitemOpen
  \bibfield  {author} {\bibinfo {author} {\bibfnamefont {W.~A.}\ \bibnamefont
  {Harrison}},\ }\href@noop {} {\emph {\bibinfo {title} {Electronic Structure
  and the Properties of Solids}}}\ (\bibinfo  {publisher} {W. H. Freeman},\
  \bibinfo {address} {San Fransisco},\ \bibinfo {year} {1980})\BibitemShut
  {NoStop}%
\bibitem [{\citenamefont {Slater}\ and\ \citenamefont {Koster}(1954)}]{slater}%
  \BibitemOpen
  \bibfield  {author} {\bibinfo {author} {\bibfnamefont {J.~C.}\ \bibnamefont
  {Slater}}\ and\ \bibinfo {author} {\bibfnamefont {G.~F.}\ \bibnamefont
  {Koster}},\ }\href {\doibase 10.1103/PhysRev.94.1498} {\bibfield  {journal}
  {\bibinfo  {journal} {Phys. Rev.}\ }\textbf {\bibinfo {volume} {94}},\
  \bibinfo {pages} {1498} (\bibinfo {year} {1954})}\BibitemShut {NoStop}%
\bibitem [{\citenamefont {Dagotto}(2002)}]{Dagotto:Book}%
  \BibitemOpen
  \bibfield  {author} {\bibinfo {author} {\bibfnamefont {E.}~\bibnamefont
  {Dagotto}},\ }\href@noop {} {\emph {\bibinfo {title} {Nanoscale Phase
  Separation and Colossal Magnetoresistance}}}\ (\bibinfo  {publisher} {Berlin:
  Springer},\ \bibinfo {year} {2002})\BibitemShut {NoStop}%
\bibitem [{\citenamefont {Lanczos}(1950)}]{Lanczos1950}%
  \BibitemOpen
  \bibfield  {author} {\bibinfo {author} {\bibfnamefont {C.}~\bibnamefont
  {Lanczos}},\ }\href@noop {} {\bibfield  {journal} {\bibinfo  {journal}
  {Journal Of Research Of The National Bureau Of Standards}\ }\textbf {\bibinfo
  {volume} {45}},\ \bibinfo {pages} {255} (\bibinfo {year} {1950})}\BibitemShut
  {NoStop}%
\bibitem [{\citenamefont {Lanczos}(1952)}]{Lanczos1952}%
  \BibitemOpen
  \bibfield  {author} {\bibinfo {author} {\bibfnamefont {C.}~\bibnamefont
  {Lanczos}},\ }\href@noop {} {\bibfield  {journal} {\bibinfo  {journal}
  {Journal Of Research Of The National Bureau Of Standards}\ }\textbf {\bibinfo
  {volume} {49}},\ \bibinfo {pages} {33} (\bibinfo {year} {1952})}\BibitemShut
  {NoStop}%
\bibitem [{\citenamefont {Wen}\ and\ \citenamefont {Niu}(1990)}]{Wen1990}%
  \BibitemOpen
  \bibfield  {author} {\bibinfo {author} {\bibfnamefont {X.~G.}\ \bibnamefont
  {Wen}}\ and\ \bibinfo {author} {\bibfnamefont {Q.}~\bibnamefont {Niu}},\
  }\href
  {\doibase 10.1103/PhysRevB.41.9377}{\bibfield  {journal} {\bibinfo  {journal} {Phys. Rev. 
  B}\ }\textbf {\bibinfo {volume} {41}},\ \bibinfo {pages} {9377} (\bibinfo
  {year} {1990})}\BibitemShut {NoStop}%
\bibitem [{\citenamefont {Tao}\ and\ \citenamefont {Wu}(1984)}]{Tao1984}%
  \BibitemOpen
  \bibfield  {author} {\bibinfo {author} {\bibfnamefont {R.}~\bibnamefont
  {Tao}}\ and\ \bibinfo {author} {\bibfnamefont {Y.~S.}\ \bibnamefont {Wu}},\
  }\href
  {\doibase 10.1103/PhysRevB.30.1097}{\bibfield  {journal} {\bibinfo  {journal} {Phys. Rev. 
  B}\ }\textbf {\bibinfo {volume} {30}},\ \bibinfo {pages} {1097} (\bibinfo
  {year} {1984})}\BibitemShut {NoStop}%
\bibitem [{\citenamefont {He}\ \emph {et~al.}(1992)\citenamefont {He},
  \citenamefont {Xie},\ and\ \citenamefont {Zhang}}]{He1992}%
  \BibitemOpen
  \bibfield  {author} {\bibinfo {author} {\bibfnamefont {S.}~\bibnamefont
  {He}}, \bibinfo {author} {\bibfnamefont {X.~C.}\ \bibnamefont {Xie}}, \ and\
  \bibinfo {author} {\bibfnamefont {F.-C.}\ \bibnamefont {Zhang}},\ }\href
  {\doibase 10.1103/PhysRevLett.68.3460}{\bibfield  {journal} {\bibinfo  {journal} {Phys. Rev. Lett.}\
  }\textbf {\bibinfo {volume} {68}},\ \bibinfo {pages} {3460} (\bibinfo {year}
  {1992})}\BibitemShut {NoStop}%
\bibitem [{\citenamefont {Johnson}\ and\ \citenamefont
  {Canright}(1994)}]{Johnson1994}%
  \BibitemOpen
  \bibfield  {author} {\bibinfo {author} {\bibfnamefont {M.~D.}\ \bibnamefont
  {Johnson}}\ and\ \bibinfo {author} {\bibfnamefont {G.~S.}\ \bibnamefont
  {Canright}},\ }\href
  {\doibase 10.1103/PhysRevB.49.2947}{\bibfield  {journal} {\bibinfo  {journal}
  {Phys. Rev.~B}\ }\textbf {\bibinfo {volume} {49}},\ \bibinfo {pages}
  {2947} (\bibinfo {year} {1994})}\BibitemShut {NoStop}%
\bibitem [{\citenamefont {Jain}(1989)}]{Jain1989}%
  \BibitemOpen
  \bibfield  {author} {\bibinfo {author} {\bibfnamefont {J.~K.}\ \bibnamefont
  {Jain}},\ }\href
  {\doibase 10.1103/PhysRevLett.63.199}{\bibfield  {journal} {\bibinfo  {journal}
  {Phys. Rev. Lett.}\ }\textbf {\bibinfo {volume} {63}},\ \bibinfo {pages} {199}
  (\bibinfo {year} {1989})}\BibitemShut {NoStop}%
\bibitem [{\citenamefont {Niu}\ \emph {et~al.}(1985)\citenamefont {Niu},
  \citenamefont {Thouless},\ and\ \citenamefont {Wu}}]{Niu1985}%
  \BibitemOpen
  \bibfield  {author} {\bibinfo {author} {\bibfnamefont {Q.}~\bibnamefont
  {Niu}}, \bibinfo {author} {\bibfnamefont {D.~J.}\ \bibnamefont {Thouless}}, \ and\ \bibinfo {author} {\bibfnamefont {Y.-S.}\ \bibnamefont
  {Wu}},\ }\href
  {\doibase 10.1103/PhysRevB.31.3372}{\bibfield  {journal} {\bibinfo  {journal}
  {Phys. Rev.~B}\ }\textbf {\bibinfo {volume} {31}},\ \bibinfo {pages}
  {3372} (\bibinfo {year} {1985})}\BibitemShut {NoStop}%
\bibitem [{\citenamefont {Xiao}\ \emph {et~al.}(2010)\citenamefont {Xiao},
  \citenamefont {Chang},\ and\ \citenamefont {Niu}}]{Xiao2010}%
  \BibitemOpen
  \bibfield  {author} {\bibinfo {author} {\bibfnamefont {D.}~\bibnamefont
  {Xiao}}, \bibinfo {author} {\bibfnamefont {M.~C.}\ \bibnamefont {Chang}}, \
  and\ \bibinfo {author} {\bibfnamefont {Q.}~\bibnamefont {Niu}},\ }\href
  {\doibase 10.1103/RevModPhys.82.1959} {\bibfield  {journal} {\bibinfo
  {journal} {Rev. Mod. Phys.}\ }\textbf {\bibinfo {volume} {82}},\
  \bibinfo {pages} {1959} (\bibinfo {year} {2010})}\BibitemShut {NoStop}%
\bibitem [{\citenamefont {Sheng}\ \emph {et~al.}(2003)\citenamefont {Sheng},
  \citenamefont {Wan}, \citenamefont {Rezayi}, \citenamefont {Yang},
  \citenamefont {Bhatt},\ and\ \citenamefont {Haldane}}]{Sheng2003}%
  \BibitemOpen
  \bibfield  {author} {\bibinfo {author} {\bibfnamefont {D.~N.}\ \bibnamefont
  {Sheng}}, \bibinfo {author} {\bibfnamefont {X.}~\bibnamefont {Wan}}, \bibinfo
  {author} {\bibfnamefont {E.~H.}\ \bibnamefont {Rezayi}}, \bibinfo {author}
  {\bibfnamefont {K.}~\bibnamefont {Yang}}, \bibinfo {author} {\bibfnamefont
  {R.~N.}\ \bibnamefont {Bhatt}}, \ and\ \bibinfo {author} {\bibfnamefont
  {F.~D.~M.}\ \bibnamefont {Haldane}},\ }\href {\doibase
  10.1103/PhysRevLett.90.256802} {\bibfield  {journal} {\bibinfo  {journal}
  {Phys. Rev. Lett.}\ }\textbf {\bibinfo {volume} {90}},\ \bibinfo
  {pages} {256802} (\bibinfo {year} {2003})}\BibitemShut {NoStop}%
\bibitem [{\citenamefont {Yang}\ \emph
  {et~al.}(2012{\natexlab{b}})\citenamefont {Yang}, \citenamefont {Sun},\ and\
  \citenamefont {{Das Sarma}}}]{Yang2012}%
  \BibitemOpen
  \bibfield  {author} {\bibinfo {author} {\bibfnamefont {S.}~\bibnamefont
  {Yang}}, \bibinfo {author} {\bibfnamefont {K.}~\bibnamefont {Sun}}, \ and\
  \bibinfo {author} {\bibfnamefont {S.}~\bibnamefont {{Das Sarma}}},\ }\href
  {http://link.aps.org/doi/10.1103/PhysRevB.85.205124} {\bibfield  {journal}
  {\bibinfo  {journal} {Phys. Rev.~B}\ }\textbf {\bibinfo {volume} {85}},\
  \bibinfo {pages} {205124} (\bibinfo {year} {2012}{\natexlab{b}})}\BibitemShut
  {NoStop}%
\end{thebibliography}

\end{document}